\documentclass[]{article}

\usepackage{graphicx}
\usepackage{epstopdf, epsfig}
\usepackage{amssymb,amsmath}
\usepackage{tikz}
\usepackage{graphicx}
\usepackage{dsfont}
\usepackage{accents}
\usepackage{pgfplots}
\usepackage{hyperref}
\usepackage{arxiv,epsf}
\usepackage{caption, subcaption}
\usepackage[export]{adjustbox}
\usepackage{float}
\usepackage{arxiv}
\usepackage{orcidlink}
\pgfplotsset{compat=1.15}
\newlength{\dhatheight}

\usepackage{newtxtext}
\usepackage{newtxmath}
\usepackage{natbib}
\usepackage{hyperref}
\hypersetup{
	colorlinks = true,
	urlcolor   = blue,
	citecolor  = black,
}

\newcommand{\RomanNumeralCaps}[1]
\linenumbers

\renewcommand{\exp}{\mathrm{e}}
\newcommand{\ci}{\mathrm{i}}
\newcommand{\im}{\mathrm{i}}
\newcommand{\wrt}{\mathrm{d}}
\newcommand{\phin}{{\hat{\psi}}}

\numberwithin{equation}{section}
\title{On the attenuation of waves through broken ice of randomly-varying thickness on water of finite depth}
\author{Lloyd Dafydd \& Richard Porter}

\begin{document}

\maketitle

\begin{abstract}
The recent work of \citet{dafydd2024attenuation} on the attenuation of waves 
propagating through floating broken ice of random thickness is 
extended to consider water of non-shallow depth. A theoretical model of 
broken floating ice is analysed using a multiple scales analysis to 
provide an explicit expression for the attenuation of waves as they 
propagate from a region of constant thickness ice into a semi-infinite 
region of ice whose thickness is a slowly-varying random function of 
distance.  Theoretical predictions are shown to compare well to 
numerical simulations of scattering over long finite regions of ice 
of randomly-varying thickness computed from an approximate 
depth-averaged model derived under a mild-slope assumption. The theory 
predicts a low-frequency attenuation proportional to the eighth power of 
frequency and a roll-over effect at higher frequencies. The relationship
between the results and field measurements are discussed.
\end{abstract}


\section{Introduction}

In a recent paper, \citet{dafydd2024attenuation} 
considered how waves attenuate
as they pass through regions of broken ice floating on water. Their study was 
conducted under a shallow water assumption allowing a basic model to be
developed which took the form of a second order ODE whose coefficients
encoded the variation of the ice thickness. Subsequently, a multiple scales
analysis, inspired by the approach adopted by \citet{grataloup2003localization}, 
\citet{mei2004evolution} and \citet{mei2005theory}, but incorporating significant 
adjustments to the methodology, was used to predict the 
decay of wave energy through a region of randomly-varying ice thickness 
as a function of distance. Theoretical predictions were shown to 
compare well with numerical simulations.

The model of \citet{dafydd2024attenuation} is energy conserving with no
source of physical damping incorporated into the dynamics. The mechanism 
responsible for the decay of wave energy propagating through a random
environment results from multiple scattering and is related to
the phenomenon of localisation, an effect first described by \citet{anderson1958absence} in quantum systems.

The main purpose of the work of \citet{dafydd2024attenuation} and of the present 
paper is to consider whether randomness and multiple scattering is a 
candidate for describing the data from field 
measurements of attenuation of wave energy collected in polar regions.
This study adds to a number of similar studies which share the same
purpose in which randomness is introduced in a number of different ways.
In \citet{kohout2008elastic}, \citet{squire2009ocean} and 
\citet{mokus2023breaking} two-dimensional scattering is considered through
floes whose length is the order of the wavelength and
where randomness is introduced through the length of the floes. In
\citet{bennetts2010three}, \citet{montiel2016attenuation}, and \citet{meylan2021floe}
three-dimensional studies were conducted into random configurations 
of randomly-sized circular floes. Many researchers have also considered 
different mechanisms by which attenuation occurs as a result of 
physical damping such as viscoelastic effects (\citet{wang2010visco}, \citet{keller1998visco}), roughness, and
a range of phenomological effects (as in \citet{vaughan2009decay}).
None of these studies have 
been able to provide a definitive account of the observed data.

A number of field studies over many decades 
including \citet{wadhams1988attenuation}, \citet{liu1992wave}, 
\citet{kohout2014storm}, \citet{doble2015relating}, 
\citet{kohout2020observations}, \citet{montiel2022physical}
have sought to establish an increasingly refined 
description of the attenuation with a particular emphasis on characterising
its dependence on wave frequency, $\omega$. For low to mid frequencies 
the data is suggestive of a power-law dependence of $\omega^{n}$ where $n$ 
lies between 2 and 4 (see, for example, \citet{meylan2014in}, 
\citet{meylan2018dispersion}, \citet{montiel2022physical}). However the 
evidence for extending the power-law fit to very low frequencies appears 
much less clear and \citet{meylan2021floe} have suggested that
models with $n=8-10$ may capture the behaviour. In fact, 
the work of \citet{herman2024apparent} suggests that the $2<n<4$ power-law dependence may be a flawed approach to evaluating models that address ice-related attenuation and in fact models with a relatively high $n$ should not necessarily be disregarded.
For higher frequencies most data sets available indicate 
a roll-over effect (see \citet{li2017rollover} for a comprehensive 
discussion) although a more recent study 
by \citet{thomson2021spurious} provides good evidence to suggest roll over
is a statistical effect due to the methods used for sampling noisy data.
The analysis of \citet{rogers2021estimates} and 
\citet{montiel2022physical} uses this work to filter out these effects.

In \citet{dafydd2024attenuation} the 
basic shallow water model, theory and numerical simulations resulted in an 
attenuation coefficient which behaved as $\omega^2$ for low frequencies 
with a roll-over effect -- a peak in attenuation followed by exponential 
decay -- at higher frequencies. \citet{dafydd2024attenuation} assumed floes of 
small size and introduced randomness in a different way to previous
authors via the ice thickness. A number of 
simplifying modelling assumptions were introduced in order to reduce the 
complexity of
a fluid interacting with broken ice to a 2nd order ordinary differential
equation which could be subjected to analysis. Since the shallow water 
assumption was the most unrealistic modelling assumption, here we 
have extended that work and report on the case of water of finite depth
albeit still working in a two-dimensional setting.

In this paper we assume an ideal fluid undergoing small amplitude motion
and hence described by a potential satisfying Laplace's 
equation.
Nevertheless, the multiple-scales analysis applies in a similar way to
\citet{dafydd2024attenuation}, with 
our methodology following \citet{mei2005theory} 
and \citet{bennetts2015absence} who investigated wave localisation over variable bathymetry in the 
presence of an unloaded surface. Significantly we modify the approach to 
incorporate the steps made in \citet{dafydd2024attenuation} which corrects an
overprediction of the attenuation. Numerical simulations of \citet{bennetts2015absence} highlighted the differences between the 
decay from averaging individual realisations of wave scattering over variable 
bathymetry with the theoretical predictions based on the method of \citet{mei2005theory} and attributed this to coherent phase cancellation in 
the averaging process in the multiple scales analysis. In \citet{dafydd2024attenuation} two modifications were made to address this: (i) a scattering problem was
formulated from the outset with incident waves entering a 
semi-infinite region of randomness; (ii) in the averaging process, terms
that could be identified as contributing to decay via coherent phase 
cancellation rather than by multiple scattering were removed from the
computation of the attenuation coefficient. The resulting numerical
simulations showed a similar character to those of \citet{bennetts2015absence} 
and were represented accurately by the theory.

In Section \ref{section2} we formulate a continuum model for floating broken ice
based on the same assumptions employed in \citet{dafydd2024attenuation}. Thus
ice completely covers the surface of the fluid and is constrained to move
vertically and independently of its neighbours. The floes are sufficiently
narrow in width for the governing equations and boundary conditions to be 
transformed into a continuous description of the ice, equivalent to the
mass loading model used by, for example, \citet{keller1953reflection}, \citet{mosig2017water} but modified for 
variable ice thickness, equivalent to the model derived by \citet{porter2004approximations} without bending forces. In Section \ref{section3}, a multiple-scales analysis is 
employed to the two-dimensional boundary-value problem the result
of which is an explicit expression for the attenuation coefficient.
In Section \ref{section4} we derive a mild-slope equation for broken ice (MSEBI) on the
assumption that the ice thickness is a slowly-varying function of
space starting from a variational principle for the governing equations.
Results of numerical solutions to the MSEBI are compared with the theory
in Section \ref{section5}. Section \ref{section5} also includes a 
discussion regarding the results of the model and how it might relate to
to data from field measurements. Further potential work and limitations 
will also be presented.

% ...existing code continues unchanged...

\section{Modelling floating broken ice}\label{section2}
	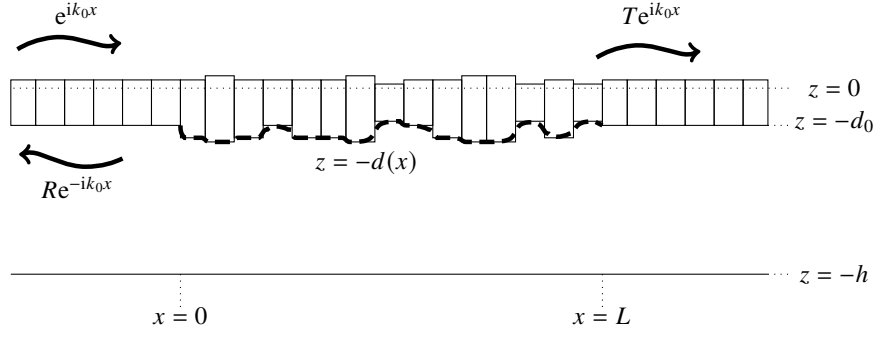
\begin{figure}[h]
	\centering
	\def \globalscale {0.550000}
	\begin{tikzpicture}[y=1cm, x=1cm, yscale=\globalscale,xscale=\globalscale, every node/.append style={scale=\globalscale}, inner sep=0pt, outer sep=0pt]
		\path[draw=black,dotted] (19.2, 25.4) -- (19.7, 25.4);
		
		\path[draw=black,dotted] (19.2, 21.8) -- (19.7, 21.8);
		
		\path[draw=black,dotted] (19.7, 26.3) -- (0.9, 26.3);
		
		\node at (2.5,23.9) {\LARGE$R\exp^{-\im k_0 x}$};
		\node at (2.5,28.1) {\LARGE$\exp^{\im k_0 x}$};
		\node at (16.4,28.1) {\LARGE$T \exp^{\im k_0 x}$};
		\node at (9.5,24.5) {\LARGE$z=-d(x)$};
		\node at (15.2,20.8) {\LARGE$x=L$};
		\node at (5.0,20.8) {\LARGE$x=0$};
		\node at (20.8,26.3) {\LARGE$z=0$};
		\node at (20.8,25.5) {\LARGE$z=-d_0$};
		\node at (20.8,21.8) {\LARGE$z=-h$};
		
		\path[draw=black,->,line cap=butt,line join=miter,line width=0.06cm,cm={ 
			-1.0,-0.1,0.1,-1.0,(2.2, 52.2)}] (3.6, 24.6).. controls (3.4, 24.5) and (3.1, 
		24.4) .. (2.7, 24.4).. controls (2.3, 24.4) and (1.9, 24.6) .. (1.6, 24.7).. 
		controls (1.3, 24.8) and (1.2, 24.7) .. (1.1, 24.7);

		\path[draw=black,->,line cap=butt,line join=miter,line width=0.06cm,cm={ 
			-1.0,-0.1,0.1,-1.0,(16.2, 52.0)}] (3.6, 24.6).. controls (3.4, 24.5) and (3.1,
		24.4) .. (2.7, 24.4).. controls (2.3, 24.4) and (1.9, 24.6) .. (1.6, 24.7).. 
		controls (1.3, 24.8) and (1.2, 24.7) .. (1.1, 24.7);

		\path[draw=black,->,line cap=butt,line join=miter,line width=0.06cm] (3.6, 24.6)..
		controls (3.4, 24.5) and (3.1, 24.4) .. (2.7, 24.4).. controls (2.3, 24.4) 
		and (1.9, 24.6) .. (1.6, 24.7).. controls (1.3, 24.8) and (1.2, 24.7) .. (1.1,
		24.7);

		\path[draw=black,dotted] (15.2, 21.8) -- (15.2, 21.0);

		\path[draw=black,dotted] (5.0, 21.8) -- (5.0, 21.0);

		\path[draw=black] (0.9, 21.8) 
		-- (19.2, 21.8);

		\path[draw=black] (18.6, 26.5) rectangle 
		(19.2, 25.4);

		\path[draw=black] (17.9, 26.5) rectangle 
		(18.6, 25.4);

		\path[draw=black] (17.2, 26.5) rectangle 
		(17.9, 25.4);

		\path[draw=black] (16.5, 26.5) rectangle 
		(17.2, 25.4);

		\path[draw=black] (15.8, 26.5) rectangle 
		(16.5, 25.4);

		\path[draw=black] (15.2, 26.5) rectangle 
		(15.8, 25.4);

		\path[draw=black] (14.5, 26.4) rectangle 
		(15.2, 25.5);

		\path[draw=black] (13.8, 26.5) rectangle 
		(14.5, 25.1);

		\path[draw=black] (13.1, 26.4) rectangle 
		(13.8, 25.5);

		\path[draw=black] (12.4, 26.6) rectangle 
		(13.1, 25.0);

		\path[draw=black] (11.8, 26.6) rectangle 
		(12.4, 25.0);

		\path[draw=black] (11.1, 26.5) rectangle 
		(11.8, 25.1);

		\path[draw=black] (10.4, 26.5) rectangle 
		(11.1, 25.4);

		\path[draw=black] (9.0, 26.6) rectangle (9.7,
		25.0);

		\path[draw=black] (8.4, 26.5) rectangle (9.0,
		25.1);

		\path[draw=black] (7.7, 26.5) rectangle (8.4,
		25.1);

		\path[draw=black] (9.7, 26.4) rectangle 
		(10.4, 25.5);

		\path[draw=black] (7.0, 26.5) rectangle (7.7,
		25.4);

		\path[draw=black] (6.3, 26.5) rectangle (7.0,
		25.1);

		\path[draw=black] (5.6, 26.6) rectangle (6.3,
		25.0);

		\path[draw=black] (5.0, 26.5) rectangle (5.6,
		25.1);

		\path[draw=black] (4.3, 26.5) rectangle (5.0,
		25.4);

		\path[draw=black] (3.6, 26.5) rectangle (4.3,
		25.4);

		\path[draw=black] (2.9, 26.5) rectangle (3.6,
		25.4);

		\path[draw=black] (2.2, 26.5) rectangle (2.9,
		25.4);

		\path[draw=black] (1.5, 26.5) rectangle (2.2,
		25.4);

		\path[draw=black] (0.9, 26.5) rectangle (1.5,
		25.4);
		
		\path[draw=black,line cap=butt,line join=miter,line width=0.06cm,miter 
		limit=4.0,dash pattern=on 0.2cm off 0.1cm] (5.0, 25.4).. controls (5.0, 25.3) 
		and (5.0, 25.2) .. (5.1, 25.1).. controls (5.2, 25.1) and (5.4, 25.1) .. (5.5,
		25.1).. controls (5.6, 25.1) and (5.6, 25.0) .. (5.7, 25.0).. controls (5.9, 
		25.0) and (6.1, 25.0) .. (6.2, 25.0).. controls (6.3, 25.0) and (6.3, 25.1) ..
		(6.4, 25.1).. controls (6.5, 25.1) and (6.8, 25.1) .. (6.9, 25.1).. controls 
		(7.0, 25.2) and (7.0, 25.3) .. (7.1, 25.3).. controls (7.2, 25.4) and (7.4, 
		25.4) .. (7.6, 25.3).. controls (7.7, 25.3) and (7.7, 25.3) .. (7.7, 25.2).. 
		controls (7.7, 25.2) and (7.7, 25.1) .. (7.8, 25.1).. controls (7.9, 25.1) and
		(8.1, 25.1) .. (8.4, 25.1).. controls (8.6, 25.1) and (8.8, 25.1) .. (8.9, 
		25.1).. controls (9.0, 25.1) and (9.0, 25.0) .. (9.1, 25.0).. controls (9.3, 
		25.0) and (9.5, 25.0) .. (9.6, 25.1).. controls (9.7, 25.2) and (9.7, 25.3) ..
		(9.8, 25.4).. controls (9.9, 25.5) and (10.2, 25.5) .. (10.3, 25.5).. 
		controls (10.4, 25.4) and (10.4, 25.4) .. (10.5, 25.4).. controls (10.6, 25.4)
		and (10.8, 25.4) .. (11.0, 25.3).. controls (11.1, 25.3) and (11.1, 25.2) .. 
		(11.2, 25.1).. controls (11.3, 25.1) and (11.5, 25.1) .. (11.6, 25.1).. 
		controls (11.8, 25.1) and (11.8, 25.0) .. (11.9, 25.0).. controls (12.0, 25.0)
		and (12.2, 25.0) .. (12.4, 25.0).. controls (12.6, 25.0) and (12.9, 25.0) .. 
		(13.0, 25.1).. controls (13.1, 25.2) and (13.1, 25.3) .. (13.2, 25.4).. 
		controls (13.3, 25.5) and (13.6, 25.5) .. (13.7, 25.4).. controls (13.8, 25.3)
		and (13.8, 25.2) .. (13.9, 25.2).. controls (14.0, 25.1) and (14.2, 25.1) .. 
		(14.4, 25.2).. controls (14.5, 25.2) and (14.5, 25.3) .. (14.6, 25.4).. 
		controls (14.7, 25.5) and (14.9, 25.5) .. (15.0, 25.5).. controls (15.2, 25.4)
		and (15.2, 25.4) .. (15.2, 25.4);

	\end{tikzpicture}
	\caption{Definition sketch of random broken ice over a flat bed of finite depth.}
\end{figure}

Cartesian coordinates, $(x,z)$, are used where $z$ is directed upwards 
from an origin located in the horizontal rest position of the water surface 
in the absence of an ice cover. Fluid of density $\rho$ is bounded below
by a flat rigid bed at $z = -h$ and above by a continuous ice cover,
broken into rectangular blocks/floes by narrow vertical cracks along
$x = x_i$. The depth of submergence of the floe lying between 
$x_i < x < x_{i+1}$ is $d_i$ and its width, $l_i = x_{i+1}-x_i$, is assumed 
small compared to the wavelength of waves that propagate across
the ice-covered water surface.
On account of the Archimedes principle the thickness of the 
ice floe in $x_i < x < x_{i+1}$ is given by
$d_i\rho/\rho_{ice}$, where $\rho_{ice}$ is the density of ice. 
	
Assuming inviscid incompressible irrotational flow, the velocity of the 
fluid is given by the gradient of a potential $\Phi(x,z,t)$. 
Small amplitude motion is assumed implying that the
governing equations and boundary conditions are linearised. We factorise a
harmonic time dependence of angular frequency $\omega$ from the motion
such that $\Phi(x,z,t) = \mathrm{Re}\left(\phi(x,z)\exp^{-\im \omega t}\right)$
where $\phi(x,z)$ is a complex potential satisfying 
\begin{equation}
 \nabla^2\phi = 0 \qquad \mbox{on $-h<z<-d_{pc}(x)$}
 \label{2.1}
\end{equation}
in the fluid where $d_{pc}(x) = d_i$ for $x_i < x < x_{i+1}$
is a piecewise constant function.
On a flat bed the impermeability condition is expressed by
\begin{equation}
 \phi_z = 0 \qquad \mbox{on $z=-h$}.
 \label{2.2}
\end{equation}
At the fluid-ice interface the kinematic boundary condition is expressed
by
\begin{equation}
 \phi_z = -\im \omega \zeta_i \qquad \mbox{on $z=-d_i$, $x_i < x < x_{i+1}$},
 \label{2.3}
\end{equation}
where $\mathrm{Re} \{ \zeta_i \exp^{-\im \omega t} \}$ is the heave
amplitude of the ice block with left- and right-hand edges along $x = x_i$
and $x=x_{i+1}$. This function is also coupled to $\phi$ through 
a dynamic condition for the constrained heave motion of each
floe by
\begin{equation}
 -\omega^2\rho \ell_i d_i \zeta_i = - \rho g \ell_i \zeta_i + \im \omega \rho \int_{x_i}^{x_{i+1}} \phi(x,-d_i)\; \wrt x \quad \mbox{on $x_i < x < x_{i+1}$}
 \label{2.4}
\end{equation}
where $g$ is gravity.

Additionally there is no flow across the vertical faces of the steps in the 
ice in contact with the fluid and this is expressed by
\begin{equation}
 \phi_x = 0, \qquad x=x_i,~ -\mbox{max}(d_{i-1},d_i) < z < -\mbox{min}(d_{i-1},d_i).
 \label{2.5}
\end{equation}
We non-dimensionalise these governing equations by writing 
\begin{equation}
 x' = Kx, \qquad
 z' = Kz, \qquad
 \phi(x,z) = -(\ci A g/\omega) \phi'(x',z'), \qquad
 \zeta_i(x) = A \zeta_i'(x'), \qquad
 \label{2.6}
\end{equation}
where $A$ is some characteristic amplitude of motion of the ice 
and $K = \omega^2/g$. This results in (after dropping primes)
\begin{equation}
 \nabla^2 \phi = 0 \qquad \mbox{in $-h < z < -d_{pc}(x)$}
 \label{2.7}
\end{equation}
with
\begin{equation}
 \phi_z = 0, \qquad \mbox{on $z = -h$}
 \label{2.8}
\end{equation}
and
\begin{equation}
 \phi_z = \zeta_i, \qquad \mbox{on $z = -d_i$, $x_i < x < x_{i+1}$}
 \label{2.9}
\end{equation}
with
\begin{equation}
 (1 - d_i) \zeta_i = \frac{1}{l_i} \int_{x_i}^{x_{i+1}} \phi(x,-d_i)
 \, dx, \qquad \qquad \mbox{on $x_i < x < x_{i+1}$}.
 \label{2.10}
\end{equation}
and $h$, $x_i$ and $d_i$ are all now dimensionless variables and $l_i
\ll 1$ for each $i$. The condition (\ref{2.8}) remains
unchanged after non-dimensionalisation.

The next step is to replace the discrete description of the ice 
with a continuum model. In $x_i < x < x_{i+1}$ we write
\begin{equation}
 l_i \zeta_i = \int_{x_i}^{x_{i+1}} \phi_z(x,-d_i) \, dx =
 \int_{x_i}^{x_{i+1}} \phi_n(x,-d(x)) \sqrt{1 + [d'(x)]^2} \, dx
 \label{2.11}
\end{equation}
using the divergence theorem in the small region $-d(x) < z < -d_{pc}(x)$ 
between the discrete
description of the ice cover and a continuous version formed by 
$z = -d(x)$ where $d(x)$ is defined such that 
$d(x_i) = \max \{d_i, d_{i-1} \}$ and $d'(x)$ exists everywhere
(practically, we define $d(x)$ and infer $d_{pc}(x)$). In deriving 
(\ref{2.11}) we have used (\ref{2.5}), (\ref{2.9}) and (\ref{2.1}) 
in the fluid region $-d(x) < z < -d_{pc}(x)$;
the operator $\partial_n = (1+[d'(x)]^2)^{-1/2}(d'(x) \partial_x + \partial_z)$ is the derivative normal to the
curve $z = -d(x)$ directed out of the fluid. Thus
\begin{equation}
 \zeta_i = \frac{1}{l_i}  \int_{x_i}^{x_{i+1}} 
 (d'(x) \phi_x(x,-d(x)) + \phi_z(x,-d(x))) \, dx  \approx d'(\bar{x}_i) \phi_x(\bar{x}_i,-d(\bar{x}_i)) + \phi_z(\bar{x}_i,-d(\bar{x}_i))
 \label{2.12}
\end{equation}
for $\bar{x}_i = \frac12 (x_i+ x_{i+1})$ with error $O(l_i^2)$.
Combining with (\ref{2.10}) and making approximations consistent with those 
made above results in
\begin{equation}
 (1 - d(x))\phi_n(x,-d(x)) = \phi(x,-d(x)).
 \label{2.13}
\end{equation}
The governing equation in the fluid is
\begin{equation}
 \nabla^2 \phi = 0 \qquad \mbox{on $-h < z < -d(x)$}
 \label{2.14}
\end{equation}
and holds in the domain bounded above by the continuous function $d(x)$
Thus (\ref{2.14}), (\ref{2.13}) and (\ref{2.8}) represents the continuum 
model governing equations for floating ice.
We note that it coincides with the mass-loading model for floating ice 
in the case $d(x)$ is a constant, assumed by \citet{weitz1950reflection}, 
\citet{keller1953reflection}, \citet{wadhams1991waves}, \citet{squire2009ocean}, \citet{mosig2017water}, etc.

We are concerned with the solution of a scattering problem where
waves are incident from $x < 0$ through a region of broken ice 
of constant submergence $d(x) = d_0$ into a region in $x > 0$
where the ice thickness/submergence is varying randomly. In this
case we pose far field boundary conditions
\begin{equation}
 \phi(x,z) \sim \left\{ 
 \begin{array}{l} (\exp^{\im k_0 x} + R_{\infty} \exp^{-\im k_0 x}) Z_0(z), \qquad x \to -\infty, \\
 0, \qquad x \to \infty \end{array} \right\}
 \label{2.15}
\end{equation}
where it is assumed that randomness gives rise to wave attenuation and 
where $R_\infty$ is the reflection coefficient satisfying 
$|R_\infty| = 1$ is required to conserve energy whilst
\begin{equation}
 Z_0(z) = N_0^{-1/2} \cosh k_0 (z+h), \qquad \mbox{where }
 N_0 = \frac12 \left( 1 + \frac{\sinh 2 k_0 (h-d_0)}{2 k_0 (h-d_0)} \right)
 \label{2.16}
\end{equation}
is the depth variation associated with propagating waves which is
found by separating variables; $k$ is the positive real root of
\begin{equation}
 k_0 \tanh k_0(h-d_0) = \kappa \equiv 1/(1-d_0).
 \label{2.17}
\end{equation}
In Section \ref{section4} the extent of this region is finite and for $x > L$
(where $L \gg 1$ implies the region is large compared to the wavelength)
$d(x) = d_0$ once again and the boundary conditions at infinity
are replaced by 
\begin{equation}
 \phi(x,z) \sim \left\{ 
 \begin{array}{l} (\exp^{\im k_0 x} + R_L \exp^{-\im k_0 x}) Z_0(z), \qquad x \to -\infty, \\
 T_L \exp^{\im k_0 x}, \qquad x \to \infty \end{array} \right\}
 \label{2.18}
\end{equation}
where $R_L$, $T_L$ are reflection and transmission coefficients 
satisfying the energy relation $|R_L|^2 + |T_L|^2 = 1$.

In the region of variable ice thickness we define
\begin{equation}
 d(x) = d_0(1 + \sigma r(x))\label{2.18}
\end{equation}
where $r$ is a random function with zero mean 
and unit variance, expressed as
\begin{equation}
	\langle r(x) \rangle = 0, 
	\qquad
	\langle r^2(x) \rangle = 1,
 \label{2.19}
\end{equation}
whilst $r(x)$ is chosen to satisfy a Gaussian correlation relation
\begin{equation}
 \langle r(x) r(\check{x}) \rangle = \exp^{-(x-\check{x})^2/\Lambda^2}.
 \label{2.20}
\end{equation}
We assume $\sigma \ll 1$ as the RMS of the vertical variations of $d(x)$
with respect to $d_0$ and $\Lambda = O(1)$ ensures that
significant variations in $x$ are on the scale of the wavelength.
We also ensure that the $r(0) = r'(0) = 0$ so that the ice thickness 
joins the constant values in $x < 0$ smoothly. 
The same applies in $x > L$ by also letting $r(L) =r'(L) = 0$.

\section{A multiple scales expansion to determine attenuation coefficient}\label{section3}

We consider the limit $L \to \infty$.
The method used to calculate the attenuation rate is based on the
the approach documented in \citet{mei2005theory} and
altered to address random variations in ice thickness over a constant
depth fluid in place of random variations in depth beneath a fluid 
free surface. Additionally, we adopt the variation in the approach
developed in \citet{dafydd2024attenuation} which corrects for the overprediction 
in attenuation rate, as identified in \citet{bennetts2015absence}, by 
considering waves entering a semi-infinite region of randomness and 
eliminating any terms contributing to wave attenuation that are not 
associated with multiple scattering.

We introduce a slow variable $X = \sigma^2 x$ 
where $\sigma \ll 1$ and write
	\begin{equation}
		\phi(x,z) = \phi_0(x,X,z) + \sigma \phi_1(x,X,z) + \sigma^2 \phi_2(x,X,z) + \ldots
 \label{3.1}
	\end{equation}
Consequently, $\partial_x \to \partial_x + \sigma^2 \partial_X$ and so
(\ref{2.14}) is
	\begin{equation}
	\left( {\nabla}^2 + 2 \sigma^2 \frac{\partial^2}{\partial x \partial X} + \sigma^4 \frac{\partial^2}{\partial X^2} \right)\left(\phi_0 + \sigma \phi_1 + \sigma^2 \phi_2 + \ldots \right)
  =0. 
 \label{3.2}
	\end{equation}
The bed condition (\ref{2.8}) becomes
	\begin{equation}
		\frac{\partial}{\partial z} \left(\phi_0 + \sigma \phi_1 + \sigma^2 \phi_2 + \ldots \right)
  = 0,\quad\text{on}\quad z = -h
 \label{3.3}
	\end{equation}
and the condition (\ref{2.13}) is expanded as
	\begin{align}
	(1- d_0 - d_0 \sigma r(x)) \left(
 \frac{\partial}{\partial z} - \sigma d_0 r(x) 
 \frac{\partial^2}{\partial z^2} + \sigma^2 \frac{d_0^2 r^2(x)}{2}
 \frac{\partial^3}{\partial z^3} + 
 \sigma d_0 r'(x) \frac{\partial}{\partial x} - \sigma^2 d_0^2 r(x) r'(x)
 \frac{\partial^2}{\partial x \partial z} + \ldots \right)
\\ \left(\phi_0 + \sigma \phi_1 + \sigma^2 \phi_2 + \ldots \right) ~~~~
 \\
 = \left( 1 + \sigma^2 \frac{d_0^2 [r'(x)]^2}{2} + \ldots \right)
 \left( 1 - \sigma d_0 r(x) 
 \frac{\partial}{\partial z} + \sigma^2 d_0^2 r^2(x) 
 \frac{\partial^2}{\partial z^2} + \ldots \right) \left(\phi_0 + \sigma \phi_1 + \sigma^2 \phi_2 + \ldots \right)
 \label{3.4}
	\end{align}
and evaluated on $z=-d_0$. We assume that the terms of $O(l_i^2)$ neglected
in formulating the continuum description of the floating ice are smaller
than $O(\sigma^2)$. Thus, the theory described below formally applies
for sufficiently narrow ice floes.

We are also required to match to (\ref{2.15}) in $x < 0$ which results in
\begin{equation}
 \phi_0(x,0,z) \sim (\exp^{\im k_0 x} + R_\infty \exp^{-\im k_0 x}) Z_0(z), \qquad
 \frac{\partial}{\partial x} \phi_0(x,0,z) \sim \im k_0 (\exp^{\im k_0 x} - R_\infty \exp^{-\im k_0 x})Z_0(z), \qquad x \to -\infty
 \label{3.5}
\end{equation}
whilst $\phi_1(x,0,z)$, $\phi_2(x,0,z), \ldots$ should behave as outgoing waves 
into $x < 0$ (i.e. proportional to $\exp^{-\im k_0 x}$ as $x \to -\infty$). As $x \to \infty$ we require $\phi_i \to 0$, $i=0,1,2,\ldots$.

\subsection{Order $1$}

At leading order we have
\begin{equation}
 \nabla^2 \phi_0 = 0, \qquad \mbox{in $-h < z < -d_0$}
 \label{3.6}
\end{equation}
with
\begin{equation}
 \frac{\partial \phi_0}{\partial z}  = 0, \qquad \mbox{on $z=-h$}
 \label{3.7}
\end{equation}
and
\begin{equation}
 \frac{\partial \phi_0}{\partial z} = \kappa \phi_0, \qquad \mbox{on $z=-d_0$}.
 \label{3.8}
\end{equation}
Solutions in $x > 0$ are therefore given by
\begin{equation}
 \phi_0 = \left( A(X) \exp^{\ci k_0 x} + B(X) \exp^{-\ci k_0 x} \right) Z_0(z)
 \label{3.9}
\end{equation}
where $k$, $Z_0$ are defined by (\ref{2.17}), (\ref{2.16}) and 
where $A$ and $B$, representing the macroscale modulation to the wave 
amplitude, are to be found. 
We also have from (\ref{2.15}),
\begin{equation}
 1 + R_\infty = A(0) + B(0), \qquad
 \im k_0(1 - R_\infty) = \im k_0(A(0) - B(0))
 \label{3.10}
\end{equation}
and so $B(0) = R_\infty$ and $A(0) = 1$.

Following the arguments in 
\citet{dafydd2024attenuation}, $|R_\infty| = 1$ requires there to be a 
local balance of energy $|A(X)| = |B(X)|$ and so $B(X)$ only varies
from $A(X)$ by a phase factor. This means the second term in (\ref{3.9})
can be written as the complex conjugate of the first term, modulo a 
phase. In order to simplify the presentation henceforth we 
present the calculations with the first term and it is to be
assumed that the second term should be added to this.

\subsection{Order $\sigma$}

At next order we have
\begin{equation}
 \nabla^2 \phi_1 = 0, \qquad \mbox{in $-h < z < -d_0$}
 \label{3.11}
\end{equation}
with
\begin{equation}
 \frac{\partial \phi_1}{\partial z} = 0, \qquad \mbox{on $z=-h$}
 \label{3.12}
\end{equation}
and
\begin{equation}
 \frac{\partial \phi_1}{\partial z} - \kappa \phi_1 = 
 d_0 r(x) \frac{\partial^2 \phi_0}{\partial z^2} 
 -
 d_0 r'(x) \frac{\partial \phi_0}{\partial x} 
, \qquad \mbox{on $z=-d_0$}.
 \label{3.13}
\end{equation}
Using (\ref{3.9}) this last equation can be rewritten as
\begin{equation}
 \frac{\partial \phi_1}{\partial z} - \kappa \phi_1 
 = d_0 Z_0(-d_0) \left(r(x)k_0^2 - \im k_0 r'(x) \right) A(X) \exp^{\im k_0 x} 
 \label{3.14}
\end{equation}
(recalling that we are henceforth suppressing the term proportional to $B(X)$).
Additionally, $\phi_1(x,0,z)$ is outgoing as $x \to -\infty$ where 
$r(x) = 0$ and $\phi_1 \to 0$ as $x,X \to \infty$ for $-h < z < d_0$.
The solution to the problem above is formulated using a Green's function 
$G(x,\check{x},z)$ defined in $-\infty < x, \check{x} < \infty$ and $-h < z< -d_0$ satisfying
\begin{align}
		&\frac{\partial^2 G}{\partial x^2}+\frac{\partial^2 G}{\partial z^2} = 0, \qquad \mbox{in $-h < z < -d_0$,} \label{3.15}\\
		&\frac{\partial G}{\partial z} = 0, \qquad \mbox{on $z=-h$, and} \label{3.16}\\
		&\frac{\partial G}{\partial z} - \kappa G = \delta(x-\check{x}), \qquad \mbox{on $z=-d_0$.}\label{3.17}
\end{align}
We also impose a radiation condition on $G$ that waves are outgoing from the
source, $x = \check{x}$.
Employing Green's Identity with $\phi_1$ and $G(x;\check{x},z)$ we have
\begin{equation}
 0 = \int_{0}^{\infty}\int_{-h}^{-d_0} \phi_1(x,X,z)\nabla^2 G(x;\check{x},z) - G(x;\check{x},z)\nabla^2\phi_1(x,X,z)\;\wrt z \; \wrt x = \int_{\partial D} \phi_1\frac{\partial G}{\partial n} - G\frac{\partial \phi_1}{\partial n} \;\wrt S
 \label{3.18}
\end{equation}
where $\partial D$ is the boundary of the area defined in the left-hand integral. Using the conditions of the problem for $\phi_1$ and those placed upon $G$ results in the left-hand side of (\ref{3.18}) being zero and only the line $z=-d_0$, $0 < x < \infty$ contributes from the boundary integral so that
\begin{equation}
	\phi_1(\check{x},X,z) = d_0 Z_0(-d_0) A(X) \int_{0}^{\infty} 
 G(x,\check{x},z)\left( k_0^2 r(x) - \im k_0 r'(x) \right)
 \exp^{\im k_0 x}\; \wrt x.
 \label{3.19}
\end{equation}

Using Fourier transforms (see e.g. \citet{mei2005theory}), it is 
readily found that 
\begin{equation}
	G(x,\check{x},z) = \frac{1}{2\pi}\int_{\Gamma}\frac{\cosh\left[\alpha(h+z)\right]\exp^{\im \alpha (x-\check{x})}}{\alpha\sinh\left[\alpha(h- d_0)\right] - \kappa\cosh\left[\alpha(h- d_0)\right]}\;\wrt\alpha
 \label{3.20}
\end{equation}
where $\Gamma$ represents the contour along the real axis of the complex $\alpha$-plane with indents below the pole at $\alpha = k_0$ and above the pole at $\alpha=-k_0$. By residue calculus, it can be shown that, along $z=- d_0$,
\begin{equation}
G(x,\check{x},-d_0) \equiv {\mathcal G}(x - \check{x}) = \frac{\im Z_0^2(-d_0) \exp^{\im k_0 |x - \check{x}|}}{2k(h-d_0)} + \sum_{n=1}^\infty \frac{Z_n^2(-d_0) \exp^{- \gamma_n |x - \check{x}|}}{2 \gamma_n(h- d_0)}
 \label{3.21}
\end{equation}
where $\alpha = \pm \ci \gamma_n$ are the location of poles on the imaginary axis and $\gamma_n$ denotes the positive real roots of
\begin{equation}
 \kappa = -\gamma_n\tan[\gamma_n(h- d_0)].
 \label{3.22}
\end{equation}
Thus, using the symmetry in $x, \check{x}$ of the Green's function, the solution to (\ref{2.16}) on the boundary $z=-d_0$ is given by
\begin{equation}
 \phi_1(x,X,-d_0) = d_0 Z_0(-d_0)A(X)\int_{0}^{\infty} {\mathcal G}(x-\check{x})\left( k_0^2 r(\check{x}) - \im k_0 r'(\check{x}) \right) \exp^{\im k_0 \check{x}}\; \wrt \check{x}.
 \label{3.23}
\end{equation}

\subsection{Order $\sigma^2$}

The governing equations for the second-order potential $\phi_2$ are 
\begin{equation}
 		\frac{\partial^2 \phi_2}{\partial x^2}+\frac{\partial^2 \phi_2}{\partial z^2} = - 2\frac{\partial^2 \phi_0}{\partial x \partial X},
\label{3.24}
\end{equation}
\begin{equation}
\frac{\partial \phi_2}{\partial z} = 0, \qquad \mbox{on $z=-h$},\label{3.25}
\end{equation}
and
\begin{multline}
	\frac{\partial \phi_2}{\partial z} - \kappa \phi_2 = 
 d_0 r(x) \frac{\partial^2 \phi_1}{\partial z^2} -d_0 r'(x) 
 \frac{\partial \phi_1}{\partial x} 
 -\frac12 d_0^ 2r^2 (x)\left( \kappa\frac{\partial^2\phi_0}{\partial z^2}  +\frac{\partial^3 \phi_0}{\partial z^3} \right) + \frac12  \kappa d_0^ 2[r' (x)]^2 \phi_0
 \\
 +d_0^2 r(x) r'(x) \left(\kappa\frac{\partial\phi_0}{\partial x}  +\frac{\partial^2 \phi_0}{\partial z \partial x}\right),
		\qquad \mbox{on $z=- d_0$}. 
 \label{3.26}
\end{multline}
We can use (\ref{3.9}), (\ref{3.11}), (\ref{3.12}) to rewrite (\ref{3.26}) as
\begin{equation}
 \frac{\partial \phi_2}{\partial z} - \kappa \phi_2 = 
 -d_0 r(x) \frac{\partial^2 \phi_1}{\partial x^2} 
 -d_0 r'(x) \frac{\partial \phi_1}{\partial x} 
 +\kappa d_0^2 k_0^2 r^2 (x) \phi_0  + \frac12  \kappa d_0^2 [r' (x)]^2 \phi_0
 +2 \im \kappa d_0^2 k r(x) r'(x) \phi_0,
 \qquad \mbox{on $z=- d_0$}.
 \label{3.27}
\end{equation}
Next we  ensemble average the governing equations for $\phi_2$ using 
(\ref{3.9}) to give
\begin{equation}
 \frac{\partial^2 \langle \phi_2 \rangle}{\partial x^2}+\frac{\partial^2 \langle \phi_2 \rangle}{\partial z^2} = - 2 \im k_0 A'(X) \exp^{\im k_0 x} Z_0(z), \qquad -h < z < -d_0
\label{3.28}
\end{equation}
and
\begin{equation}
 \frac{\partial \langle \phi_2 \rangle}{\partial z} = 0, \qquad \mbox{on $z=-h$},\label{3.29}
\end{equation}
and
\begin{equation}
 \frac{\partial \langle \phi_2 \rangle}{\partial z} - \kappa \langle \phi_2 \rangle = 
 -d_0 \left\langle r(x) \frac{\partial^2 \phi_1}{\partial x^2} \right\rangle 
 -d_0 \left\langle r'(x) 
 \frac{\partial \phi_1}{\partial x} \right\rangle
 + d_0^2 \kappa (k_0^2 +\Lambda^{-2}) A(X) \exp^{\im k_0 x} Z_0(-d_0),
		\qquad \mbox{on $z=- d_0$},
 \label{3.30}
\end{equation}
using (\ref{2.19}) and noting that $\langle r(x) r'(x) \rangle = 0$ and
$\langle r'(x)^2 \rangle = 2/\Lambda^2$. The first two
terms on the right-hand side need to be considered  carefully before addressing 
the problem for $\langle \phi_2 \rangle$ further. Thus, we first note
from (\ref{3.23}) that
\begin{equation}
 \frac{\partial  \phi_1}{\partial x} 
 = -d_0 Z_0(-d_0) A(X) \int_0^\infty \hspace{-17pt} - \hspace{12pt}
 \frac{\partial {\mathcal G}}{\partial \check{x}} \left( k_0^2 r(\check{x}) - \im k_0 r'(\check{x}) \right)
 \exp^{\im k_0 \check{x}} \; d \check{x}
 \label{3.31}
\end{equation}
since $\partial_x {\mathcal G} = -\partial_{\check{x}} {\mathcal G}$ is Cauchy singular and the integral 
is thus momentarily interpreted as a Cauchy-principal value integral. 
Integrating by parts and using the vanishing of $r$ and $r'$ at the 
limits of the integral gives
\begin{equation}
 \frac{\partial  \phi_1}{\partial x} 
 = d_0 Z_0(-d_0) A(X) \int_0^\infty
 {\mathcal G}(x-\check{x}) \, \exp^{\im k_0 \check{x}} \left( \frac{d}{d\check{x}} + \im k_0 \right) \left( k_0^2 r(\check{x}) - \im k_0 r'(\check{x}) \right)
  \; d \check{x}.
 \label{3.32}
\end{equation}
In exactly the same manner we find
\begin{equation}
 \frac{\partial^2 \phi_1}{\partial x^2} 
 = d_0 Z_0(-d_0) A(X) \int_0^\infty
 {\mathcal G}(x-\check{x}) \, \exp^{\im k_0 \check{x}} \left( \frac{d}{d\check{x}} + \im k_0 \right)^2 \left( k_0^2 r(\check{x}) - \im k_0 r'(\check{x}) \right)
  \; d \check{x}.
 \label{3.33}
\end{equation}
The next step is to multiply (\ref{3.32}) and (\ref{3.33}) by $r'(x)$ 
and $r(x)$, respectively, and take ensemble averages according to (\ref{2.19}) 
and (\ref{2.20}). This results in 
\begin{equation}
 \left\langle r'(x) \frac{\partial \phi_1}{\partial x}  \right\rangle
 = -d_0 Z_0(-d_0) \im k_0 A(X) \int_0^\infty
 {\mathcal G}(x-\check{x}) \, \exp^{\im k_0 \check{x}} \left( \frac{d}{d\check{x}} + \im k_0 \right)^2 \frac{d}{d x}
 \exp^{-(x-\check{x})^2/\Lambda^2} 
  \; d \check{x}
 \label{3.34}
\end{equation}
and
\begin{equation}
 \left\langle r(x) \frac{\partial^2 \phi_1}{\partial x^2}  \right\rangle
 = -d_0 Z_0(-d_0) \im k_0 A(X) \int_0^\infty
 {\mathcal G}(x-\check{x}) \, \exp^{\im k_0 \check{x}} \left( \frac{d}{d\check{x}} + \im k_0  \right)^3
 \exp^{-(x-\check{x})^2/\Lambda^2} 
  \; d \check{x}.
 \label{3.35}
\end{equation}
As described in \citet{dafydd2024attenuation}, we choose to discard contributions 
that lead  to ``fictitious decay'' through phase cancellations
in the averaging process and these are identified as the integrals
over $0 < \check{x} < x$. Thus, the remaining integrals from $\check{x} > x$
contain contributions that arise from scattering events upwave of the 
source point only, meaning that we proceed with the expressions
\begin{equation}
 \left\langle r'(x) \frac{\partial \phi_1}{\partial x}  \right\rangle
 = d_0 Z_0(-d_0) \im k_0 A(X) \int_x^\infty
 {\mathcal G}(x-\check{x}) \, \exp^{\im k_0 \check{x}} \left( \frac{d}{d\check{x}} + \im k_0 \right)^2 \frac{d}{d\check{x}}
 \exp^{-(x-\check{x})^2/\Lambda^2} 
  \; d \check{x}
 \label{3.36}
\end{equation}
and
\begin{equation}
 \left\langle r(x) \frac{\partial^2 \phi_1}{\partial x^2}  \right\rangle
 = -d_0 Z_0(-d_0) \im k_0 A(X) \int_x^\infty
 {\mathcal G}(x-\check{x}) \, \exp^{\im k_0 \check{x}} \left( \frac{d}{d\check{x}} + \im k_0  \right)^3
 \exp^{-(x-\check{x})^2/\Lambda^2} 
  \; d \check{x}.
 \label{3.37}
\end{equation}
We remark that a later consequence of this choice is that there is local 
conservation of energy, $|A(X)| = |B(X)|$, with both $A$ and $B$ decaying 
as $X \to \infty$. In (\ref{3.30}) we note the particular combination of 
(\ref{3.36}) and (\ref{3.37}) featured leads to the following simplification
\begin{equation}
 \left\langle r(x) \frac{\partial^2 \phi_1}{\partial x^2}  \right\rangle
 + 
 \left\langle r'(x) \frac{\partial \phi_1}{\partial x}  \right\rangle
 = d_0 Z_0(-d_0) k_0^2 A(X) \int_x^\infty
 {\mathcal G}(x-\check{x}) \, \exp^{\im k_0 \check{x}} \left( \frac{d}{d\check{x}} + \im k_0 \right)^2 \exp^{-(x-\check{x})^2/\Lambda^2} 
  \; d \check{x}.
 \label{3.38}
\end{equation}
Using the substitution $\xi = (\check{x} -x)/\Lambda$ allows us to 
write
\begin{equation}
 \left\langle r(x) \frac{\partial^2 \phi_1}{\partial x^2}  \right\rangle
 + 
 \left\langle r'(x) \frac{\partial \phi_1}{\partial x}  \right\rangle
 = \frac{d_0 Z_0(-d_0) k_0^2 A(X)}{\Lambda} \exp^{\im k_0 x} \int_0^\infty
 {\mathcal G}(\Lambda \xi) \, \exp^{\im k_0 \Lambda \xi} \left( \frac{d}{d \xi} +  \im k_0 \Lambda \right)^2
 \exp^{-\xi^2}
  \; d \xi.
 \label{3.39}
\end{equation}
We can see that the integral is 
independent of $x$ and the right-hand side is proportional to $\exp^{\ci k_0 x}$. 
In order to proceed, we insert the expression for ${\mathcal G}$
into the integral above. Thus, we have from (\ref{3.21})
\begin{multline}
 \int_0^\infty
 {\mathcal G}(\Lambda \xi) \, \exp^{\im k_0 \Lambda \xi} \left( \frac{d^2}{d \xi^2} + (k\Lambda)^2 \right)  
 \exp^{-\xi^2}
  \; d \xi
 = \frac{\im Z_0^2(-d_0)}{2 k (h-d_0)} 
 \int_0^\infty
 \exp^{2 \im k_0 \Lambda \xi} \,
 \left( \frac{d}{d\xi} + \im k_0 \Lambda \right)^2
 \exp^{-\xi^2} 
  \; d \xi
 \\
 + \sum_{n=1}^\infty \frac{Z_n^2(-d_0)}{2 \gamma_n (h-d_0)} 
 \int_0^\infty
 \exp^{-\gamma_n \Lambda \xi +\im k_0 \Lambda \xi} \,
 \left( \frac{d}{d\xi} +  \im k_0 \Lambda  \right)^2
 \exp^{-\xi^2}
  \; d \xi
 \label{3.40}
\end{multline}
The integrals above are calculated by integrating by parts and using 
the results found in \cite{mei2005theory} to give
\begin{equation}
 \int_0^\infty
 \exp^{2 \im k_0 \Lambda \xi} \,
 \left( \frac{d}{d\xi} + (k\Lambda) \right)^2
 \exp^{-\xi^2} 
  \; d \xi = -\frac{(k \Lambda)^2 \sqrt{\pi}}{2}
 \exp^{-k_0^2 \Lambda^2} (1+ \im \mbox{erfi}(k \Lambda))
 \label{3.41}
\end{equation}
and
\begin{equation}
 \int_0^\infty
 \exp^{-\gamma_n \Lambda \xi +\im k_0 \Lambda \xi} \,
 \left( \frac{d^2}{d\xi^2} + (k\Lambda)^2  \right)
 \exp^{-\xi^2}
  \; d \xi = -\gamma_n \Lambda - \im k_0 \Lambda 
 + (\gamma_n \Lambda)^2 \frac{\sqrt{\pi}}{2} \exp^{(\gamma_n \Lambda - \im k_0 \Lambda)^2/4} \mbox{erfc} \{ (\gamma_n \Lambda - \im k_0 \Lambda)/2 \},
 \label{3.42}
\end{equation}
where $\mbox{erfc}(\cdot)$ and $\mbox{erfi}( \cdot)$ are the 
complementary error function and the imaginary error function respectively.
Returning to the equations (\ref{3.30}), (\ref{3.39}) we now note that 
$\exp^{\im k_0 x}$ is a common factor on the right-hand side of the equations 
governing $\langle \phi_2 \rangle$. Thus we employ the ansatz
$\langle \phi_2 \rangle = \exp^{\im k_0 x}F(X,z)$ from which it follows that $F$ satisfies
\begin{equation}
 \frac{\partial^2F}{\partial z^2} - k_0^2 F = -2 \im k_0 A'(X)  Z_0(z), \qquad -h < z < -d_0
 \label{3.43}
\end{equation}
and
\begin{equation}
 \frac{\partial F}{\partial z} = 0 \quad \text{on} \quad z=-h
 \label{3.44}
\end{equation}
and
\begin{equation}
 \frac{\partial F}{\partial z} - \kappa F = 
 -d_0 \left\langle r(x) \frac{\partial^2 \phi_1}{\partial x^2} \right\rangle 
 \exp^{-\im k_0 x}
 -d_0 \left\langle r'(x) 
 \frac{\partial \phi_1}{\partial x} \right\rangle
 \exp^{-\im k_0 x}
 + d_0^2 \kappa (k_0^2 +\Lambda^{-2}) A(X) Z_0(-d_0),
 \qquad \mbox{on $z=- d_0$}.
 \label{3.45}
\end{equation}
The problem for $F(X,z)$ does not need to be solved in order to 
progress. Instead, following \citet{mei2005theory}, we apply Green's identity to $F(X,z)$ 
and $Z_0(z)$ over $-h<z<-d_0$
\begin{equation}
	\int_{-h}^{-d_0} F(X,z) Z_0''(z) - Z_0(z)\frac{\partial^2 F}{\partial z^2}(X,z)\;\wrt z = \left[ F(X,z)Z_0'(z) - Z_0(z)\frac{\partial F}{\partial z}(X,z)\right]_{-h}^{-d_0},
 \label{3.46}
\end{equation}
and it follows that
\begin{equation}
 2\im k_0 (h-d_0)  A'(X) = -Z_0(-d_0) \left( \frac{\partial F}{\partial z}(X,-d_0) - \kappa F(X,-d_0) \right)
 \label{3.47}
\end{equation}
where (\ref{3.21}), (\ref{3.22}) have been used along with 
$Z_0''(z) = k_0^2 Z_0(z)$, $Z_0'(-h) = 0$, $Z_0'(-d_0) = \kappa Z_0(-d_0)$ and
\begin{equation}
 \int_{-h}^{-d_0} Z_0^2(z) \; \wrt z = (h-d_0)
 \label{3.48}
\end{equation}
according to the definition of $Z_0(z)$ in (\ref{2.16}).
Bringing the results above together gives
\begin{equation}
 A'(X) = - \lambda A(X)
 \label{3.49}
\end{equation}
where
\begin{multline}
 \lambda = \frac{\sqrt{\pi} d_0^2 Z_0^4(-d_0) k_0^2 \Lambda \exp^{-k_0^2 \Lambda^2}(1+ \im \mbox{erfi}(k \Lambda))}{8 (h-d_0)^2} - \frac{\im d_0^2 
 Z_0^2(-d_0) (k_0^2 + \Lambda^{-2})}{2 (h-d_0) }
 \\
 + \frac{\im d_0^2 Z_0^2(-d_0) k_0^2}{4 (h-d_0)^2} 
 \sum_{n=1}^\infty \left(\frac{\sqrt{\pi}}{2} \gamma_n \Lambda
 \exp^{(\gamma_n - \im k_0)^2 \Lambda^2/4} \mbox{erfc}((\gamma_n \Lambda
 - \im k_0 \Lambda)/2) -1- \frac{\im k_0}{\gamma_n} \right) Z_n^2(-d_0).
 \label{3.50}
\end{multline}
We can confirm that the infinite series converges since 
$\gamma_n \sim n \pi/(h-d_0)$, $Z_n^2(-d_0) \sim 2$ and
\begin{equation}
 \frac{\sqrt{\pi}}{2} \gamma_n  \exp^{(\gamma_n - \im k_0)^2 \Lambda^2/4} \mbox{erfc}((\gamma_n \Lambda
 - \im k_0 \Lambda)/2) \sim 1 + \frac{\im k_0}{\gamma_n} + O(1/n^2)
 \label{3.51}
\end{equation}
as $n \to \infty$. The solution of (\ref{3.49}) with $A(0) = 1$ is given by
\begin{equation}
 A(X) = \exp^{-\lambda X}.
 \label{3.52}
\end{equation}
In parallel to the calculations we imagine having also made calculations for terms proportional to $B(X)$ whose functional dependence of $x$ is the conjugate of the terms multiplying $A(X)$. This means that 
$B(X) = R_\infty \exp^{-\lambda* X}$. Thus, the 
leading order solution is, from (\ref{3.9}) 
\begin{equation}
 \phi(x,z) \approx \phi_0(x,\sigma^2 x,z) = \exp^{-\lambda \sigma^2 x + \im k_0 x} + 
 \exp^{-\lambda^* \sigma^2 x - \im k_0 x + \ci \delta }
 \label{3.53}
\end{equation}
where $*$ denotes complex conjugation and 
since $|R_\infty| = 1$ we can write $R_\infty = \exp^{\im \delta}$. Thus, the 
theoretical prediction for the attenuation coefficient is
\begin{equation}
 \langle k_i \rangle = \sigma^2 \Re \{ \lambda \}.
 \label{3.54}
\end{equation}
The first term in (\ref{3.50}) is the dominant term in the attenuation
coefficient and originates from
multiple scattering from propagating waves; the infinite series 
in (\ref{3.50}) has a real part that also contributes to a much lesser
extent (as evidenced by numerical results) and originates from evanescent wave interactions. Thus, a reasonable approximation to make is
\begin{equation}
 \langle k_i \rangle \approx \frac{\sqrt{\pi} \sigma^2 d_0^2 Z_0^4(-d_0) 
 k_0^2 \Lambda \exp^{-k_0^2 \Lambda^2}}{8 (h-d_0)^2} 
 \label{3.55}
\end{equation}
In deep water, $(h-d_0)$ is large, $Z_0(-d_0) \sim \sqrt{2 k_0 (h-d_0)}$ and so
\begin{equation}
 \langle k_i \rangle \approx \frac{\sqrt{\pi} \sigma^2 d_0^2 
 k_0^4 \Lambda \exp^{-k_0^2 \Lambda^2}}{2}.
 \label{3.56}
\end{equation}
In shallow water, $(h-d_0)$ is small, $Z_0(-d_0) \sim 1$ and so
\begin{equation}
 \langle k_i \rangle \approx \frac{\sqrt{\pi} \sigma^2 d_0^2
 k_0^2 \Lambda \exp^{-k_0^2 \Lambda^2}}{8 (h-d_0)^2} 
 \label{3.57}
\end{equation}
and this agrees with \citet{dafydd2024attenuation}.

These formulae are unchanged upon redimensionalisation of variables. We note that since $k_0 \propto \omega$ for shallow water, $\langle \langle k_i \rangle \rangle \propto \omega^2$ for small frequencies. In deep water $k_0 \propto \omega^2$ and so $\langle \langle k_i \rangle \rangle \propto \omega^8$ for small frequencies.

In all depths, there is a peak in the value of $\langle k_i \rangle$ at intermediate values of frequency before exponential decay at higher frequencies. The peaks are at $k_0 \Lambda \approx \sqrt{2}$ and $k_0 \Lambda \approx 1$ for deep and shallow water limits respectively. The theory has been developed under the assumption that $k_0 \Lambda \not\gg 1$ and so we may expect the validy of the theoretical results to be in question for wave frequencies far beyond the peak although there is no reason to suppose that the existence of a peak attenuation frequency falls outside the modelling assumptions.

\section{A mild-slope equation for broken ice}\label{section4}

In order to test the theory developed above, we want to compare with 
numerical simulations of wave scattering through long finite regions of
broken ice of randomly-varying thickness. In pursuit of this goal, we 
set out in this section to formulate an ODE for wave scattering by
means of depth-averaging the original two-dimensional boundary-value
problem described in Section \ref{section2}.

Consider the variational principle $\delta \mathcal{L}=0$ where $\mathcal{L}$ is a functional defined by
\begin{equation}
 \mathcal{L}[\psi] = \frac{1}{2}\int_{D} \left \{ \frac{1}{1-d(x)}(\psi^2)|_{z=-d(x)} - \int_{-h}^{-d(x)} (\nabla\psi)^2 \; \wrt z \right\} \; \wrt x
 \label{4.1}
\end{equation}
where $D$ denotes the horizontal interval of the fluid/ice domain. 
Assuming variations which vanish at the boundary of the domain (or 
satisfy a radiation condition if the domain is infinite) it is readily
shown that $\mathcal{L}$ is 
stationary at $\psi = \phi$ if and only if $\phi$ satisfies the 
governing equations (\ref{2.13}), (\ref{2.14}) and (\ref{2.8}).

We can generate approximate solutions to the governing equations 
using the variational principle $\delta L = 0$. Specifically, we
choose to approximate $\psi \approx \phi$ using the standard mild-slope
ansatz that
\begin{equation}
 \psi(x,z) = \phin(x)w(x,z),
 \label{4.2}
\end{equation}
where
\begin{equation}
 w(x,z) = \frac{\cosh\left[k(h+z)\right]}{\cosh\left[k(h-d(x))\right]}
 \label{4.3}
\end{equation}
represents the depth variation associated with propagating waves
and $k = k(x) = k(d(x))$ is the real, positive root of
\begin{equation}
 k\tanh\left[k(h-d(x))\right] = \frac{1}{1 - d(x)}
 \label{4.4}
\end{equation}
being the dispersion relation for $k$ assuming $d(x)$ is locally a 
constant. The function $\phin(x)$ is determined by 
substitution of (\ref{4.2}) into (\ref{4.1}) and enforcing 
$\delta \mathcal{L} = 0$
to variations in $\phin(x)$ which vanish on the boundaries
of $D$. After integrating by parts and applying the Leibniz rule, we find 
that $\phin(x)$ is determined by solutions of
\begin{equation}
 \frac{\wrt}{\wrt x}\left(\int_{-h}^{-d(x)} w^2 \;\wrt z \frac{\wrt \phin}{\wrt x}\right) + \left(\int_{-h}^{-d(x) }w w_{zz}\;\wrt z + R(w)\right)\phin = 0,
 \label{4.5}
\end{equation}
where
\begin{equation}
 R(w) = \int_{-h}^{-d(x)} ww_{xx} \;\wrt z - \left[ w \left(w_z + d'(x) w_x - \frac{w}{1-d(x)} \right)\right]_{z=-d(x)}.
 \label{4.6}
\end{equation}
We note that
\begin{equation}
 \frac{\partial w}{\partial z} = \frac{k\sinh\left[k(h+z)\right]}{\cosh\left[k(h-d)\right]},
 \label{4.7}
\end{equation}
is zero on $z=-h$ and satisfies $w_z = w/(1-d(x))$ on $z=-d(x)$ by virtue of (\ref{4.4}). Letting
\begin{equation}
 u_0(d) = \int_{-h}^{-d(x)}w^2\;\wrt z,
 \label{4.8}
\end{equation}
and substituting into (\ref{4.5}) we find
\begin{equation}
 \frac{\wrt}{\wrt x}\left(u_0\frac{\wrt \phin}{\wrt x}\right) + \left(k^2u_0 + r(d)\right)\phin = 0
 \label{4.9}
\end{equation}
where $r = R(w)$ is defined by
\begin{equation}
 r(d) = \int_{-h}^{-d(x)} w\frac{\partial^2 w}{\partial x^2}\;\wrt z - d'(x) \left[w \frac{\partial w}{\partial x}\right]_{z=-d(x)}.
 \label{4.10}
\end{equation}
Henceforth, we will use a dot to denote differentiation with 
respect to $d$ and note from (\ref{4.4}) that
\begin{equation}
 \dot{k}
 = \frac{2k^2\cosh^2\left[k(h-d)\right]}{2k(h-d)+\sinh\left[2k(h-d)\right]}.
 \label{4.11}
\end{equation}
It follows that
\begin{equation}
 \frac{\partial w}{\partial x} = d'(x) \dot{w}
 \label{4.12}
\end{equation}
where, after some algebra, it can be found that
\begin{equation}
\dot{w} = \dot{k} \text{sech}\left[k(h-d)\right]\left((z+h)\sinh\left[k(h+z)\right]-\left((h-d)\tanh\left[k(h-d)\right]-\frac{1}{k}\right)\tanh^2\left[k(h-d)\right]\cosh\left[k(z+h)\right]\right).
 \label{4.13}
\end{equation}
Next, we have
\begin{equation}
 w\frac{\partial^2w}{\partial x^2} = d'^2(x)\ddot{w} w + d''(x) \dot{w} w.
 \label{4.14}
\end{equation}
and note that if we write
\begin{equation}
 u_1(d) = \int_{-h}^{-d} w \dot{w} \; \wrt z, \qquad u_2(d) = \dot{u}_1 - \int_{-h}^{-d} \dot{w}^2 \;\wrt z
 \label{4.15}
\end{equation}
then
\begin{equation}
	r(d) = d''(x) u_1(d) + d'^2(x) u_2 (d).
 \label{4.16}
\end{equation}
Thus, we arrive at the mild-slope equation for broken ice, 
\begin{equation}
 \frac{\wrt}{\wrt x}\left(u_0\frac{\wrt \phin}{\wrt x}\right) + \left(k^2 u_0 + d''(x) u_1(d) + d'^2(x) u_2 (d) \right)\phin = 0.
 \label{4.17}
\end{equation}
In the language of \cite{chamberlain1995modified} this version is prefixed with `Modified' since the extra terms proportional to $d'^2(x)$ and $d''(x)$ were omitted from previous versions of the mild-slope equations (in this case for variable bathmetry and an unloaded free surface). \citet{porter2020mild} later showed how a transformation of (\ref{4.17}) can produce a simplified version of the mild slope equation and we follow that approach below.

We seek to transform $\phin$ by letting
\begin{equation}
	\phin(x) = s(d)\varphi(x)
 \label{4.18}
\end{equation}
where the scaling factor $s(d(x))$ is to be determined.
We now have from (\ref{4.17}) that
\begin{equation}
	u_0 s \varphi''(x) + \left(d'(x) \left((u_0s)\dot{~} + u_0 \dot{s}\right)\right)\varphi'(x) + \left(k^2u_0s + d''(x) \tilde{u}_1+ d'^2(x) \tilde{u}_2 \right)\varphi(x) = 0
 \label{4.20}
\end{equation}
where
\begin{equation}
 \tilde{u}_1 = u_1s + u_0\dot{s}, \qquad \tilde{u}_2 = u_2 s
 + (u_0\dot{s})\dot{~}.
 \label{4.21}
\end{equation}
Following \citet{porter2020mild} we choose to remove the $d''(x)$ term by
setting $\tilde{u}_1 = 0$. This can be achieved by noting that 
$2u_0\dot{k}/k = 1$ and so
\begin{equation}
 2u_1 = \dot{u}_0 + 1 = \dot{u}_0 + 2u_0 \dot{k}/k.
 \label{4.22}
\end{equation}
Using this in (\ref{4.21}) with $\tilde{u}_1 = 0$ allows us to integrate
up to give $u_0^{1/2} k s = c$ for some arbitrary constant, $c$ which, without loss of generality, can be set equal to unity. Thus 
\begin{equation}
 s(d) = \frac{1}{k(d)\left(u_0(d)\right)^{1/2}}
 \label{4.23}
\end{equation}
defines the scaling factor that eliminates $d''(x)$ from the governing
equation and it follows that the original equation (\ref{4.17}) can be recast
as
\begin{equation}
 \left(\frac{\varphi'}{k^2}\right)' + \left(1 + \frac{\tilde{u}_2}{u_0 s k^2} d'^2(x) \right)\varphi = 0
 \label{4.24}
\end{equation}
where the original dependent variable is recovered from
\begin{equation}
 \phin(x) = \frac{\varphi(x)}{k(d)\left(u_0(d)\right)^{1/2}}.
 \label{4.25}
\end{equation}
Now, we can assume the continuous functions $d(x)$ satisfy the 
mild-slope assumption, i.e.
\begin{equation}
	\left|\frac{d'(x)}{k(h-d(x))}\right|\ll 1
 \label{4.26}
\end{equation}
which justifies the neglect of the final term in (\ref{4.24}) so that we
end up with the simpler governing equation
\begin{equation}
\left(\frac{\varphi'}{k^2}\right)' + \varphi = 0.
 \label{4.27}
\end{equation}
It is notable that this equation was derived by \citet{porter2020mild} for
scattering of surface waves over variable bathymetry and the only
difference here is that $k$ is defined by the dispersion relation 
(\ref{4.4}) for variable thickness ice on the surface over a flat bed instead of 
the dispersion relation for an unloaded surface over a variable bed.

It can be shown that
the inclusion of variable depth in addition to variable ice thickness
leads to the same equation with $k$ defined by (\ref{4.4}) but with the constant
$h$ replaced by $h(x)$.

\section{Results and discussion}\label{section5}

\begin{figure}[!h]
	\centering
	\includegraphics[width=0.5\linewidth]{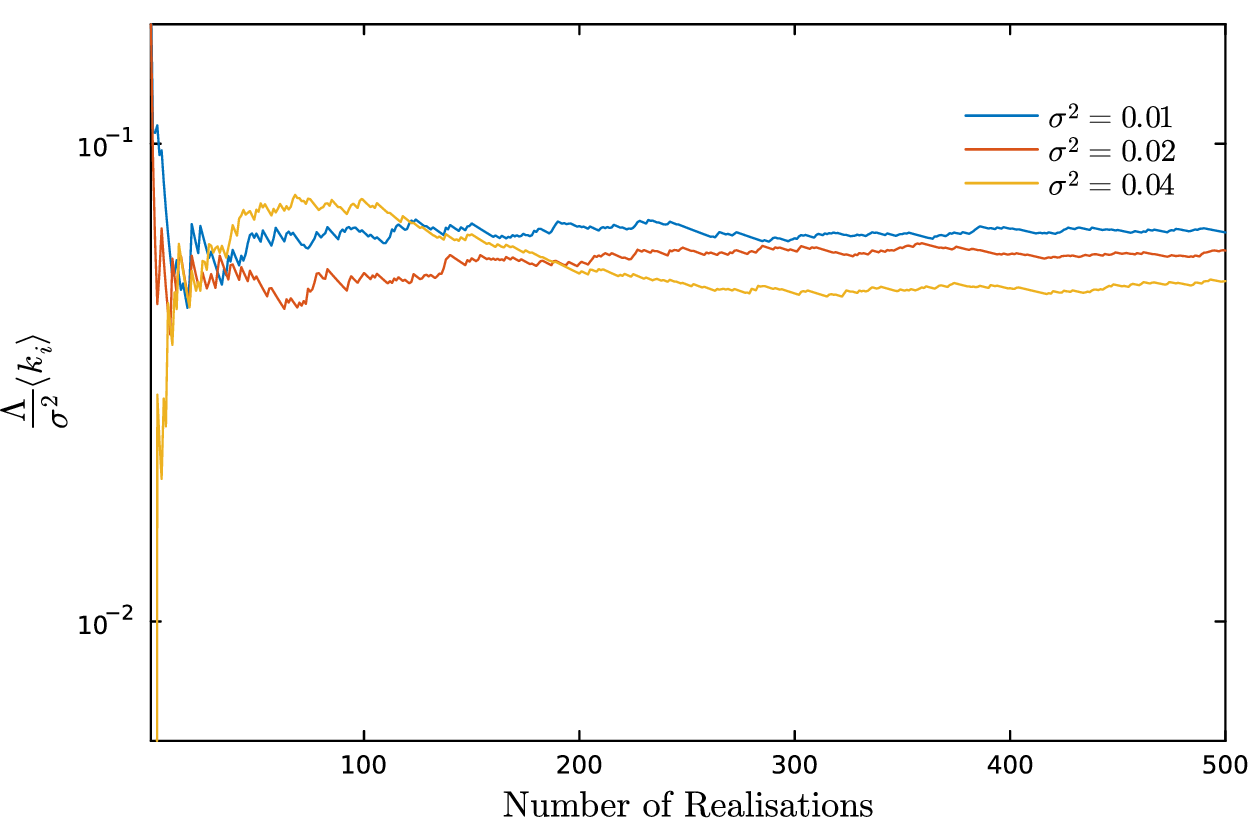}
	\caption{The convergence rate of how many random ice coverings are required for convergence for the MSEFI method.
\label{fig2}}
\end{figure}
\begin{figure}[!h]
	\centering
	\begin{subfigure}[t]{0.03\textwidth}
		\text{(a)}
	\end{subfigure}
	\begin{subfigure}[t]{0.95\textwidth}        \centering
		\includegraphics[width=\linewidth, trim=0 10 0 140, clip, valign=t]{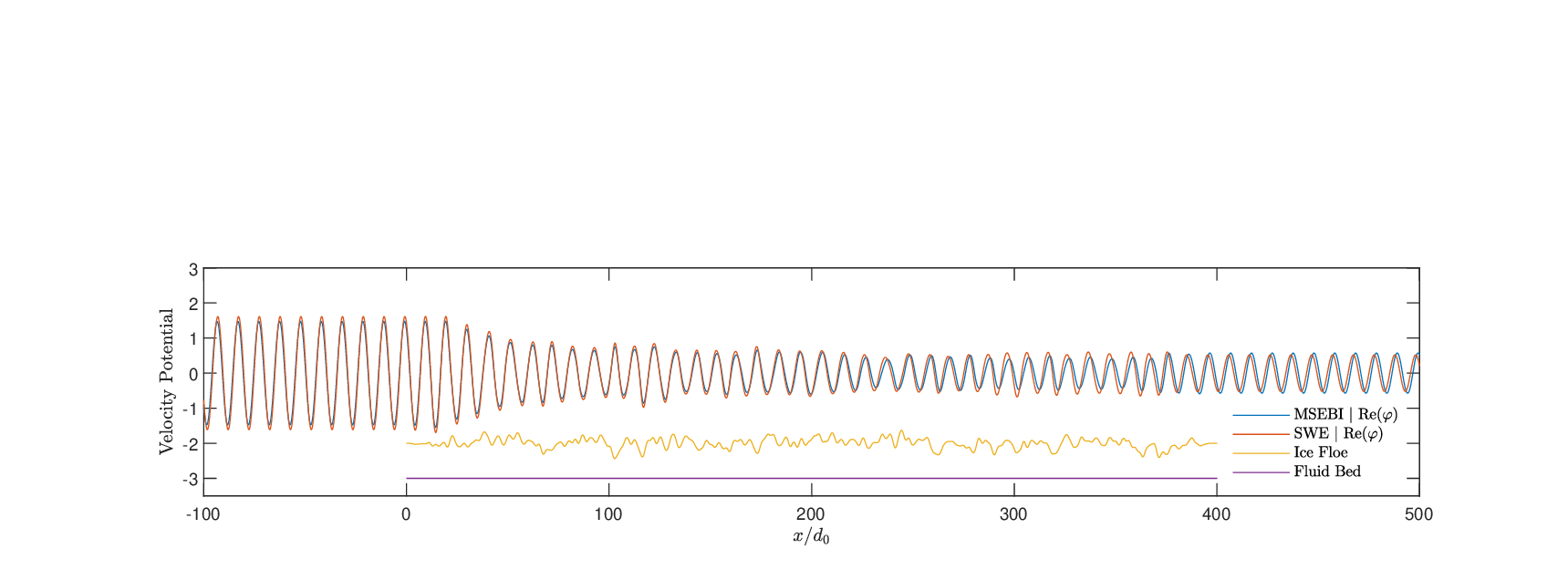}
	\end{subfigure}\hfill
	\begin{subfigure}[t]{0.03\textwidth}
		\text{(b)}
	\end{subfigure}
	\begin{subfigure}[t]{0.95\textwidth}        \centering
		\includegraphics[width=\linewidth, trim=0 10 0 140, clip, valign=t]{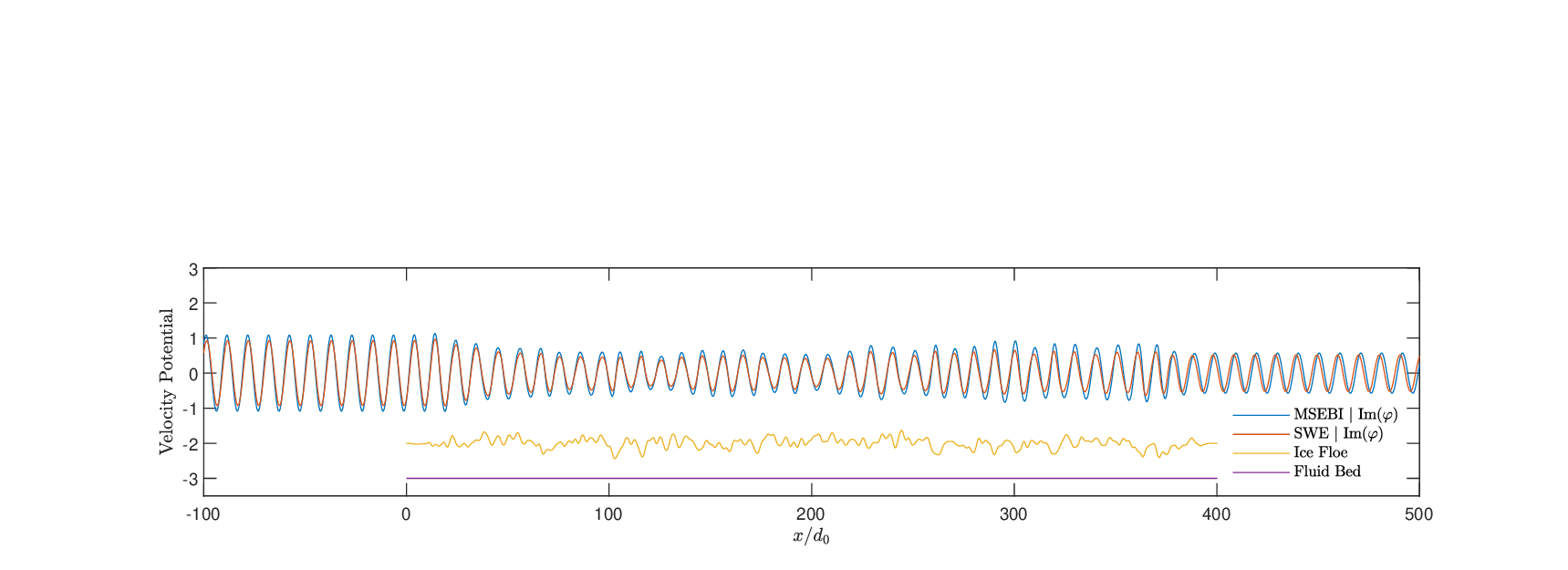}
	\end{subfigure}
	\begin{subfigure}[t]{0.03\textwidth}
		\text{(c)}
	\end{subfigure}
	\begin{subfigure}[t]{0.95\textwidth}        \centering
		\includegraphics[width=\linewidth, trim=0 10 0 140, clip, valign=t]{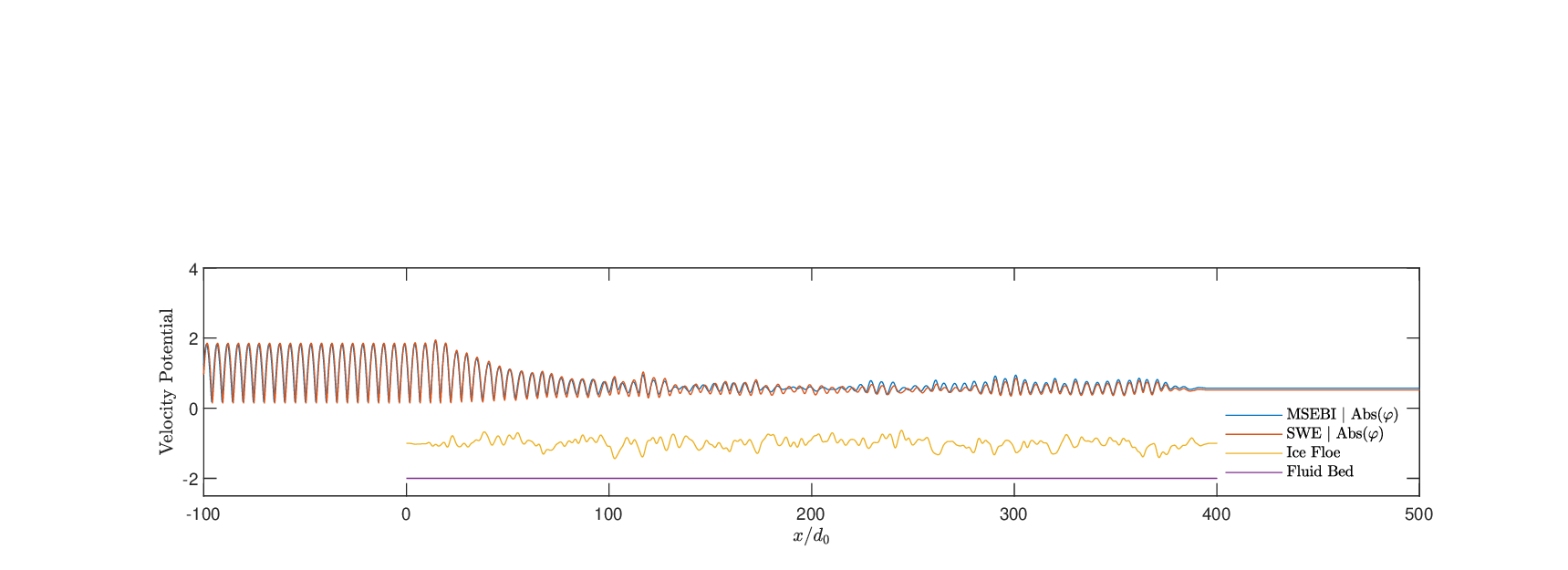}
	\end{subfigure}\hfill
	\caption{Plots comparing the MSEBI and SWE solutions for (a) real, (b) imaginary and (c) absolute values of $\varphi(x)$ over an example region of length $L/d_0 = 400$ where $\Lambda/d_0 = 2$ and $\sigma^2=0.01$.\label{fig3}}
\end{figure}
\begin{figure}[!h]
	\centering
	\begin{subfigure}[t]{0.03\textwidth}
		\text{(a)}
	\end{subfigure}
	\begin{subfigure}[t]{0.45\textwidth}        \centering
		\includegraphics[width=\linewidth, trim=0 10 0 140, clip, valign=t]{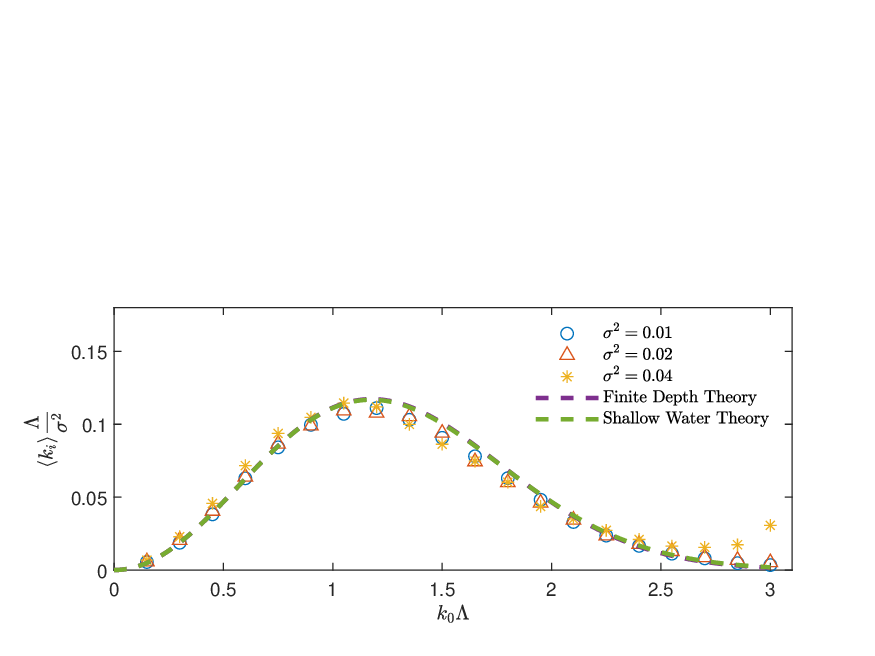}
	\end{subfigure}\hfill
	\begin{subfigure}[t]{0.03\textwidth}
		\text{(b)}
	\end{subfigure}
	\begin{subfigure}[t]{0.45\textwidth}        \centering
		\includegraphics[width=\linewidth, trim=0 10 0 140, clip, valign=t]{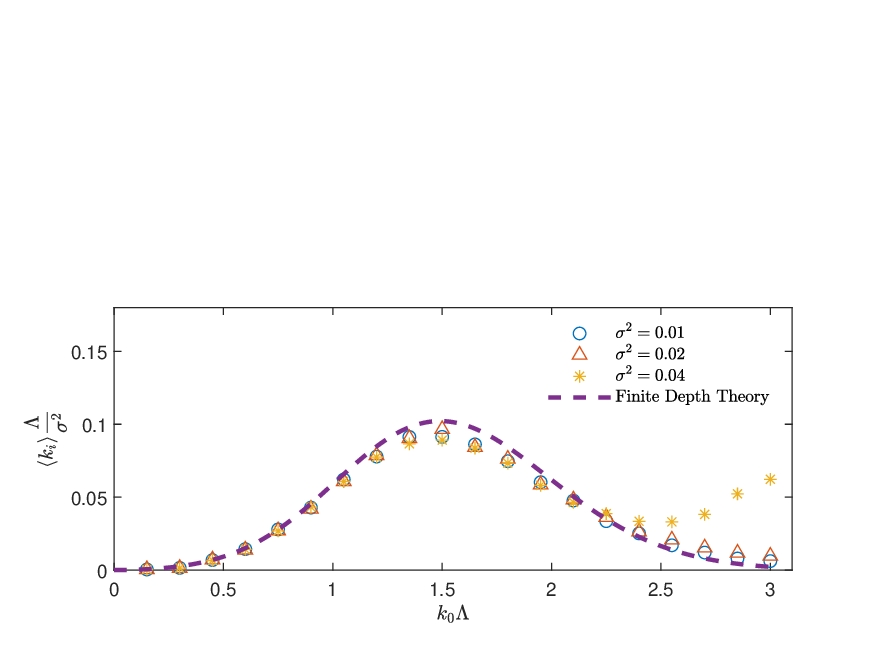}
	\end{subfigure}
	\begin{subfigure}[t]{0.03\textwidth}
		\text{(c)}
	\end{subfigure}
	\begin{subfigure}[t]{0.45\textwidth}        \centering
		\includegraphics[width=\linewidth, trim=0 10 0 140, clip, valign=t]{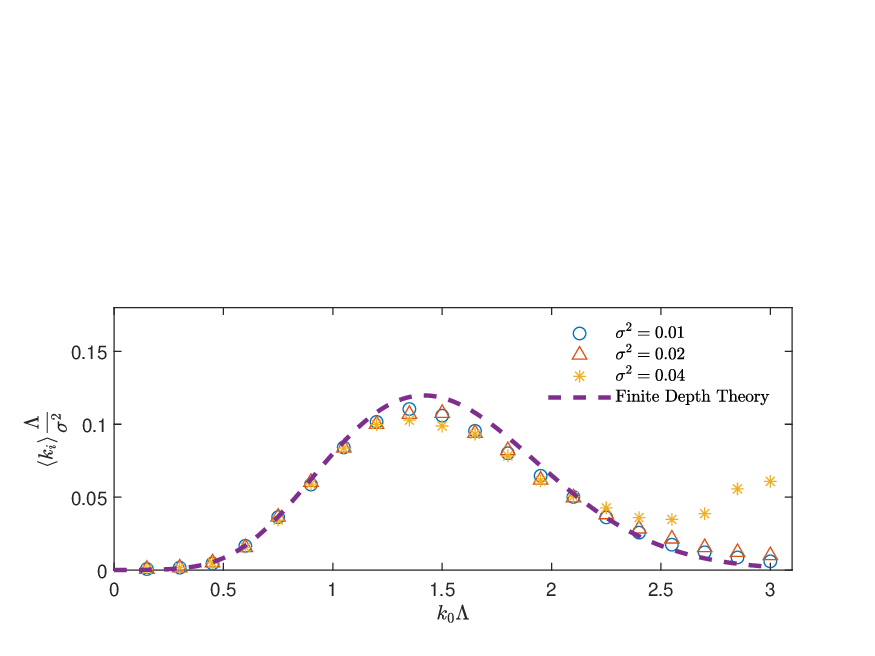}
	\end{subfigure}\hfill
	\begin{subfigure}[t]{0.03\textwidth}
		\text{(d)}
	\end{subfigure}
	\begin{subfigure}[t]{0.45\textwidth}        \centering
		\includegraphics[width=\linewidth, trim=0 10 0 140, clip, valign=t]{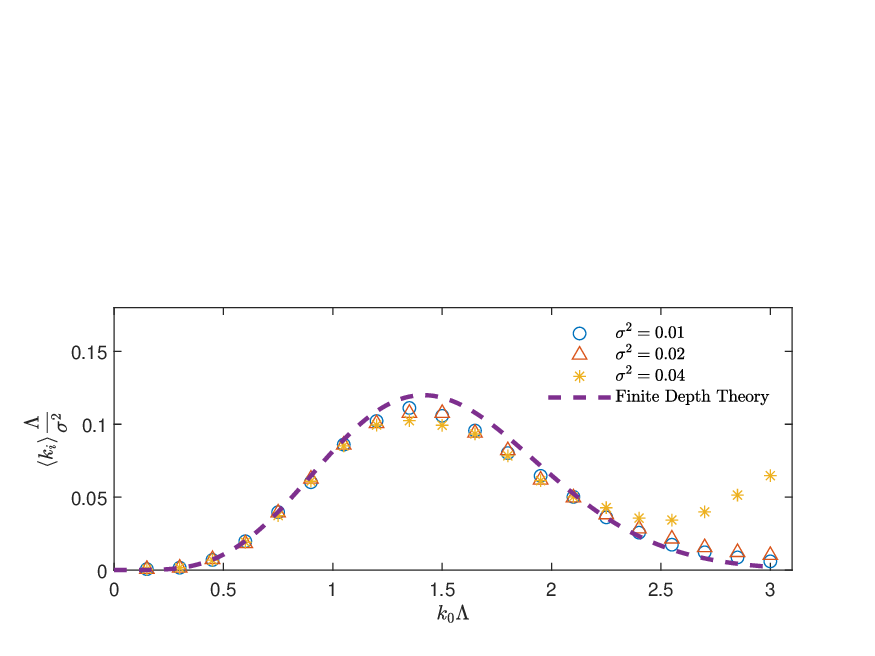}
	\end{subfigure}
	\caption{Plots comparing the derived theoretical results against the MSEFI ODE simulations for 500 realisations where $\Lambda/d_0=2$. (a) $h/d_0 = 2$, (b) $h/d_0 = 4$, (c) $h/d_0 = 8$, (d) $h/d_0 = 16$.
\label{fig4}}
\end{figure}
\begin{figure}[!h]
	\centering
	\begin{subfigure}[t]{0.03\textwidth}
		\text{(a)}
	\end{subfigure}
	\begin{subfigure}[t]{0.45\textwidth}        \centering
		\includegraphics[width=\linewidth, trim=0 10 0 140, clip, valign=t]{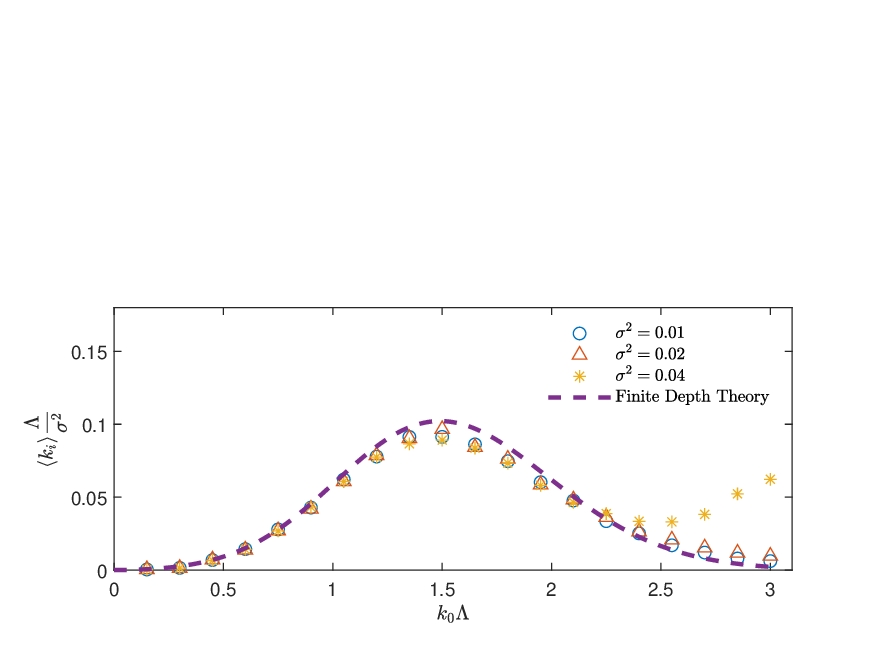}
	\end{subfigure}\hfill
	\begin{subfigure}[t]{0.03\textwidth}
		\text{(b)}
	\end{subfigure}
	\begin{subfigure}[t]{0.45\textwidth}        \centering
		\includegraphics[width=\linewidth, trim=0 10 0 140, clip, valign=t]{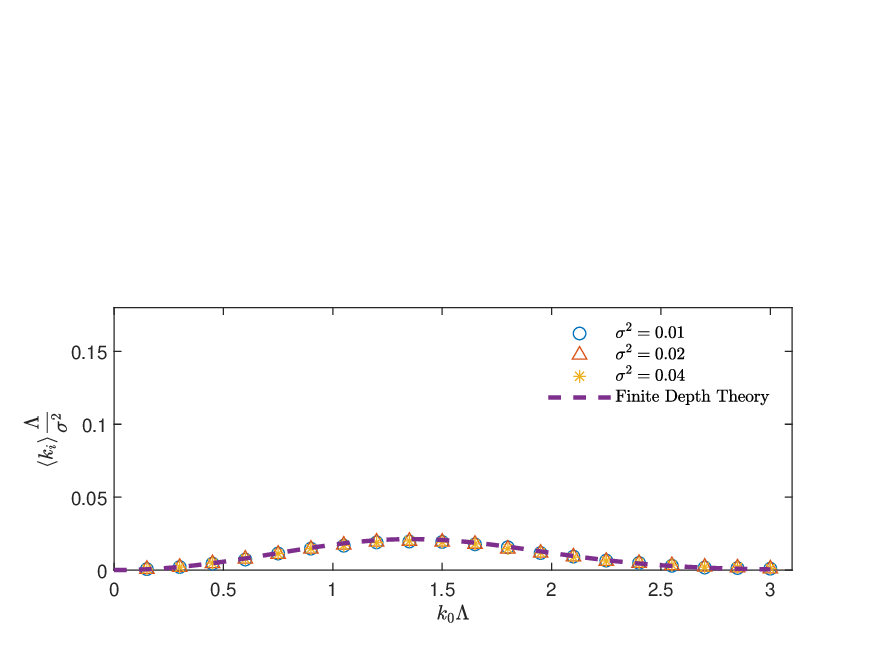}
	\end{subfigure}
	\caption{Plots comparing the derived theoretical results against the MSEFI ODE simulations for 500 realisations where $h/d_0 = 4$. (a) $\Lambda/d_0 = 2$, (b) $\Lambda/d_0 = 4$.}
\end{figure}
\begin{figure}[!h]
	\centering
	\begin{subfigure}[t]{0.03\textwidth}
		\text{(a)}
	\end{subfigure}
	\begin{subfigure}[t]{0.45\textwidth}        \centering
		\includegraphics[width=\linewidth, trim=0 10 0 140, clip, valign=t]{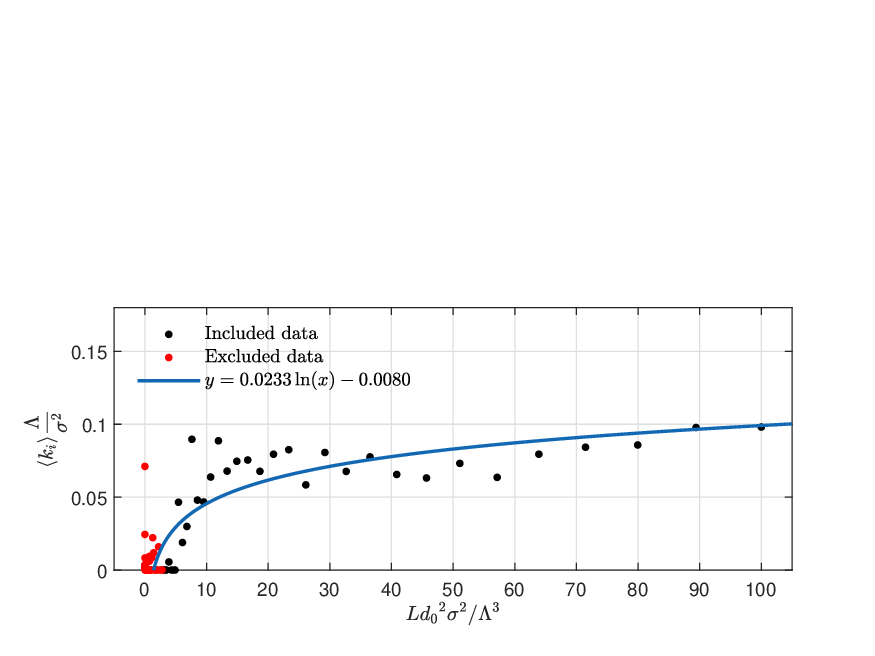}
	\end{subfigure}\hfill
	\begin{subfigure}[t]{0.03\textwidth}
		\text{(b)}
	\end{subfigure}
	\begin{subfigure}[t]{0.45\textwidth}        \centering
		\includegraphics[width=\linewidth, trim=0 10 0 140, clip, valign=t]{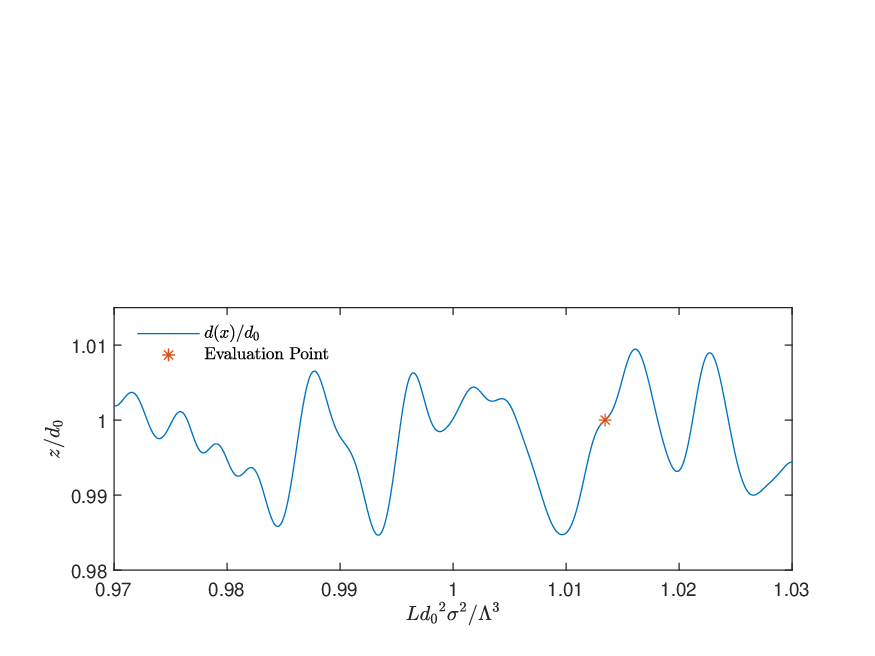}
	\end{subfigure}
	\caption{(a) Convergence of the attenuation coefficient for one region of broken ice as the length of the region ice is increased, here $k_0\Lambda = 1.5$, $h/d_0=4$. A logarithmic line of best fit is plotted. (b) A section of the random broken ice and a point of sampling where $d(x)/d_0 = 1$.}\label{figure5}
\end{figure}
The purpose of this section is to compare results from numerical 
simulations of wave scattering over long finite regions of 
randomly-varying ice thickness to the theoretical predictions of 
Section \ref{section3}. Simulations are calculated using the MSEBI described in Section \ref{section4} in which 
the problem of wave scattering is reduced through depth averaging to
solving an ordinary differential equation (ODE)
based on the assumption that $d(x)$ is a slowly-varying function.
The computation of solutions to the ODE is described in \citet{dafydd2024attenuation}.

For the simulations, $d(x)$ is taken to be a random function defined by
$d(x) = d_0(1+r(x))$ with $r(x)$ satisfying
$\langle r(x) \rangle = 0$, $\langle r(x)^2 \rangle = 1$ in $0 < x < L$ with $d(x) = d_0$ for $x < 0$ and $x > L$. The
method for defining $r(x)$ is described in \citet{dafydd2024attenuation}.
The randomness is characterised by $\sigma$ which is the RMS of 
vertical variations with respect to $d_0$ and $\Lambda$ which is the
correlation lengthscale. The attenuation is computed from the 
eigenvalues of the transfer matrix that arises from the propagation
of wave solutions over $0 < x < L$ in the same manner as described
in \citet{dafydd2024attenuation}. In order to compute approximations 
to the attenuation coefficient we must fulfil two requirements.
The first is that the value of $L$ is large enough and here
$L=20\Lambda^3/\sigma$ is chosen to anticipate the length required for convergence to be of the form $\langle k_i \rangle = C$ for some constant $C$. This choice also benefits from being frequency independent, allowing for multiple values of $k_0\Lambda$ per realisation. As in \citet{dafydd2024attenuation}, this form has the benefits of it being frequency independent and convergent at the same point 
independent of $\sigma$ and $\Lambda$ at higher frequencies. The second 
is that we average each numerically-derived $k_i$ over sufficiently large 
number, $N$, of independent random realisations of $d(x)$. 
In Figure \ref{fig2} the convergence of the average attenuation, $\langle k_i \rangle$, with increasing $N$ is shown for values up to 1500 for the MSEBI computations. The three pairs of curves show the differences from different $\sigma$ taken when $k_0 \Lambda = 1$.
On this basis we have chosen $N=500$ for 
subsequent computations.

In Figure \ref{fig4} we have plotted the dimensionless attenuation 
coefficient $\Lambda\langle k_i \rangle/\sigma^2$ against $K\Lambda$ for
$\sigma^2 = 0.01,~0.02,~0.04$. The water depth increases from shallow in subplot
(a) to deep in subplot (d). In each example, $\Lambda/d_0 = 2$ is fixed
and the data for different $\sigma^2$ should collapse onto the same
curve. 

Simulations, averaged over $N=500$ realisations, are computed using the MSEBI method. These results are shown to follow the theoretical predictions of Section \ref{section3}. The effect of adding the contribution from the infinite sum to the formula for $\langle k_i \rangle$ is virtually indistinguishable from the leading order term (\ref{3.55}). We note that the when the depth is shallow the curves in Figure \ref{fig4} are similar in character to those presented in \citet{dafydd2024attenuation}. 

For small $k_0 (h-d_0)$. and therefore small $K(h-d_0)$ where $K = \omega^2/g$, we see that the attenuation is proportional to $\omega^2$.  For larger depths, i.e. $k_0(h-d_0)\gg 1$, the low frequency behaviour
changes and the attenuation tends towards $\omega^8$.

For larger values of $\sigma^2$, the discrepancy between the theoretical predictions and numerical simulations increases at higher frequencies. This behaviour is anticipated, as the theory is formally valid only for low frequencies and small $\sigma$. Since a similar theoretical result can be obtained from the depth-averaged model in the \hyperref[AppendixA]{Appendix}, it is likely that the primary source of this discrepancy is the increased value of $\sigma$.

It should be noted that greater agreement with the theory can be achieved through increasing the number of random realisations, $N$, used in the averaging process and increasing the surface length $L$. However, this latter alteration comes at a great computational cost as due to the rapidly-varying nature of the function $d(x)$ in contrast to the slowly-varying $k(x)$ and thus a very small tolerance and step-size is required alongside the use of an implicit ODE solver. In Figure \ref{figure5}, a plot showing the convergence is shown for $k_0\Lambda = 1.5$ shows that perhaps values of $L/d_0 \geq 100\Lambda^3/\sigma^2{d_0}^3$ maybe be required to achieve high levels of convergence around the peak of the curve. This is computationally unfeasible at the precision we require.

The low frequency attenuation scaling of $\omega^8$ for deep water 
predicted in this paper is outside the range of $\omega^2$ to $\omega^4$ 
normally associated with field measurements (see \citet{meylan2018dispersion}). 
However, the work of \citet{squire2009ocean} and \citet{meylan2021floe} both suggest that at very low frequencies the attenuation data may scale 
somewhere in the range of $\omega^8$ to $\omega^{10}$.
Although the current basic model for attenuation based on two-dimensional 
multiple scattering due to randomness in the thickness of floating ice 
does not appear to fit data across all frequencies, it may still form a 
good basis for describing attenuation at low frequencies. Additionally, the work of \citet{herman2024apparent} suggests that a model's efficacy does not correspond to its ability to capture the power law of the ``apparent attenuation'' as these measurements do not adjust for energy loss that is not ice-related and in fact a higher power of $\omega$ may in fact be more desirable than the commonly desired $\omega^2$ to $\omega^{4}$. We should 
be careful not to read too much into other features of our results.
However, it is interesting to note that the present model, like that of 
\citet{dafydd2024attenuation}, exhibits a high-frequency ``roll-over'' effect 
at all depths which is a feature of field data reported in numerous
work including \citet{wadhams1988attenuation}, \citet{liu1992wave}, 
\citet{squire2009ocean}, \citet{doble2015relating}. On the other hand
the onset of roll-over in the data is itself disputed with \citet{thomson2021spurious} providing evidence that this is a statistical effect caused
by noisy data.

Within the general framework of the theory developed here, is easy to 
imagine that {\it any} model of multiple scattering through a random environment 
defined by a spatially-varying local wavenumber will result in a 
high-frequency roll-over effect. There is therefore an argument that
randomness and multiple scattering within a more complex description
of floating ice may lead to predicitions of attenuation which 
follow the field measurements.
A modification to the model presented here would imagine that individual
floes of ice are separated by small gaps of open water and, instead of
moving in constrained heave motion, are allowed to float freely.
Ongoing work by the authors have considered such a model to 
determine a local dispersion relation which assumed identical periodic
floes separated by equal gaps in terms of the ice thickness, and the
size of the gaps. 

\section{Conclusions}\label{section6}

This paper has considered a linear model for propagation of waves over 
regions of variable ice in a fluid of finite depth. A simple expression 
has been analytically derived for the ensemble averaged attenuation of
waves propagating through fragmented floating ice over a semi-infinite 
domain. In this derivation, terms associated with incoherent phase 
cancellations were removed to preserve only attenuation due to multiple 
scattering effects. The theory was shown to not only agree with the 
prior work of \citet{dafydd2024attenuation} in the shallow water regime, but 
also compares favourably with numerical simulations based on a finite-depth. The mild-slope
equation for broken ice of slowly-varying thickness was derived in the 
simple two-dimensional case over a constant dept, relevant to the present
study, but can easily be extended to three-dimensional
scattering over variable bathymetry.

The results have provided a theoretical prediction for the frequency
dependence of attenuation through broken ice which has not captured
the measurements reported in the literature across all frequencies.
However, there are some features of the results which positively
correlate to the observed data and suggest that randomness and 
multiple scattering are a dominant factor in attenuation of waves 
through regions of broken ice. It is important to note however that 
this model does not account for many of the complicated aspects of 
multi-phase, discrete sea ice. Further work such as discretising the 
blocks of sea ice and adding gaps between them to make them 
non-continuous are planned, as well as randomising gaps and 
three-dimensional scattering.

\textbf{[Funding]} {L.D. is grateful for the support of an EPSRC (UK) studentship.}\\

\textbf{[Declaration of interests]} {The authors report no conflict of interest.}\\

\textbf{[Author ORCIDs]}\\

	\orcidlink{0000-0003-1009-0946} L. Dafydd, https://orcid.org/0000-0003-1009-0946\\
	\orcidlink{0000-0003-2669-0188} R. Porter, https://orcid.org/0000-0003-2669-0188

\section*{Appendix: derivation of attenuation coefficient from the MSE for broken ice}\label{AppendixA}
\setcounter{equation}{0}
\renewcommand{\theequation}{A.\arabic{equation}}

The following section is less rigorous due to being reliant on multiple small parameters, but is included for completeness. Using (\ref{4.27}) we can trivially perform an asymptotic expansion by defining the ice submergence by the previously defined parameters (\ref{2.18}-\ref{2.20}) and then expanding our function in powers of $\sigma$ such that
	\begin{equation}
		\varphi(x) \approx \varphi_0(x,X)+ \sigma \varphi_1(x,X) + \sigma^2
		\varphi_2(x,X) + \ldots \label{A.93}
	\end{equation}
	in $x > 0$ where $X = \sigma^2 x$ is a slow variable.
	We can then rewrite (\ref{4.27}) in the form
	\begin{equation}
		((1+ \sigma C_1 r(x) + \ldots) \varphi')' + k_0^2 \varphi = 0, \qquad x > 0
		\label{A.94}
	\end{equation}
	where
	\begin{equation}
		C_1 = -\frac{2d_0k_0}{\tanh\left[k_0(h-d_0)\right]+k_0(h-d_0)\text{sech}^2\left[k_0(h-d_0)\right]}.\label{A.95}
	\end{equation}
	is found from from Taylor expanding $1/k^2(d(x))$ in $d(x)$ in powers of $\sigma$. 
	From \citet{dafydd2024attenuation} we know that (\ref{A.94}) gives us an attenuation coefficient of the form
	\begin{equation}
		\langle k_i \rangle = \frac{\sqrt{\pi}}{8} k_0^2 \sigma^2 \Lambda C_1^2 \exp^{-k_0^2 \Lambda^2}\label{A.96}
	\end{equation}
	and we can substitute our value for $C_1$ in to get
	\begin{equation}
		\langle k_i \rangle = \frac{\frac{\sqrt{\pi}}{2}{d_0}^2{k_0}^4\sigma^2\Lambda\exp^{-{k_0}^2\Lambda^2}}{\left(k_0(h-d_0)\text{sech}^2\left[k_0(h-d_0)\right]+\tanh\left[k_0(h-d_0)\right]\right)^2}.\label{A.97}
	\end{equation}
	This is identical to the formula derived for our finite depth model without the small evanescent terms.

\bibliographystyle{plainnat}
\bibliography{bibliography}

\begin{thebibliography}{34}
\providecommand{\natexlab}[1]{#1}
\providecommand{\url}[1]{\texttt{#1}}
\expandafter\ifx\csname urlstyle\endcsname\relax
  \providecommand{\doi}[1]{doi: #1}\else
  \providecommand{\doi}{doi: \begingroup \urlstyle{rm}\Url}\fi

\bibitem[Anderson(1958)]{anderson1958absence}
P.~W. Anderson.
\newblock Absence of diffusion in certain random lattices.
\newblock \emph{Phys. Rev.}, 109\penalty0 (5):\penalty0 1492--1505, 1958.
\newblock \doi{10.1103/PhysRev.109.1492}.

\bibitem[Bennetts et~al.(2010)Bennetts, Peter, Squire, and
  Meylan]{bennetts2010three}
L.~G. Bennetts, M.~A. Peter, V.~A. Squire, and M.~H. Meylan.
\newblock A three-dimensional model of wave attenuation in the marginal ice
  zone.
\newblock \emph{Journal of Geophysical Research}, 115:\penalty0 C12043, 2010.
\newblock \doi{10.1029/2009JC005982}.

\bibitem[Bennetts et~al.(2015)Bennetts, Peter, and Chung]{bennetts2015absence}
L.~G. Bennetts, M.~A. Peter, and H.~Chung.
\newblock Absence of localisation in ocean wave interactions with a rough
  seabed in intermediate water depth.
\newblock \emph{Quarterly Journal of Mechanics and Applied Mathematics},
  68\penalty0 (1):\penalty0 97--113, 2015.
\newblock \doi{10.1093/qjmam/hbu024}.

\bibitem[Chamberlain and Porter(1995)]{chamberlain1995modified}
P.~G. Chamberlain and D.~Porter.
\newblock The modified mild-slope equation.
\newblock \emph{Journal of Fluid Mechanics}, 291:\penalty0 393--407, 1995.
\newblock \doi{10.1017/S0022112095002758}.

\bibitem[Dafydd and Porter(2024)]{dafydd2024attenuation}
Lloyd Dafydd and Richard Porter.
\newblock Attenuation of long waves through regions of irregular floating ice
  and bathymetry.
\newblock \emph{Journal of Fluid Mechanics}, 996:\penalty0 A43, 2024.
\newblock \doi{10.1017/jfm.2024.655}.

\bibitem[Doble et~al.(2015)Doble, Carolis, Meylan, Bidlot, and
  Wadhams]{doble2015relating}
M.~J. Doble, G.~De Carolis, M.~H. Meylan, J.-R. Bidlot, and P.~Wadhams.
\newblock Relating wave attenuation to pancake ice thickness, using field
  measurements and model results.
\newblock \emph{Geophysical Research Letters}, 42:\penalty0 4473--4481, 2015.
\newblock \doi{10.1002/2015GL063628}.

\bibitem[Grataloup and Mei(2003)]{grataloup2003localization}
G{\'e}raldine~L. Grataloup and Chiang~C. Mei.
\newblock Localization of harmonics generated in nonlinear shallow water waves.
\newblock \emph{Phys. Rev. E}, 68\penalty0 (2):\penalty0 026314, 2003.
\newblock \doi{10.1103/PhysRevE.68.026314}.

\bibitem[Herman(2024)]{herman2024apparent}
Agnieszka Herman.
\newblock From apparent attenuation towards physics-based source terms--a
  perspective on spectral wave modeling in ice-covered seas.
\newblock \emph{Frontiers in Marine Science}, 11:\penalty0 1413116, 2024.
\newblock \doi{10.3389/fmars.2024.1413116}.

\bibitem[Keller and Weitz(1953)]{keller1953reflection}
J.~B. Keller and M.~Weitz.
\newblock Reflection and transmission coefficients for waves entering or
  leaving an icefield.
\newblock \emph{Communications on Pure and Applied Mathematics}, 6:\penalty0
  415--417, 1953.
\newblock \doi{10.1002/cpa.3160060306}.

\bibitem[Keller(1998)]{keller1998visco}
Joseph~B. Keller.
\newblock Gravity waves on ice-covered water.
\newblock \emph{Journal of Geophysical Research: Oceans}, 103\penalty0
  (C4):\penalty0 7663--7669, 1998.
\newblock \doi{10.1029/97JC02966}.

\bibitem[Kohout et~al.(2014)Kohout, Williams, Dean, et~al.]{kohout2014storm}
A.~Kohout, M.~Williams, S.~Dean, et~al.
\newblock Storm-induced sea-ice breakup and the implications for ice extent.
\newblock \emph{Nature}, 509:\penalty0 604--607, 2014.
\newblock \doi{10.1038/nature13262}.

\bibitem[Kohout and Meylan(2008)]{kohout2008elastic}
A.~L. Kohout and M.~H. Meylan.
\newblock An elastic plate model for wave attenuation and ice floe breaking in
  the marginal ice zone.
\newblock \emph{Journal of Geophysical Research}, 113:\penalty0 C09016, 2008.
\newblock \doi{10.1029/2007JC004434}.

\bibitem[Kohout et~al.(2020)Kohout, Smith, Roach, Williams, Montiel, and
  Williams]{kohout2020observations}
A.~L. Kohout, M.~Smith, L.~A. Roach, G.~Williams, F.~Montiel, and M.~J.~M.
  Williams.
\newblock Observations of exponential wave attenuation in antarctic sea ice
  during the pipers campaign.
\newblock \emph{Annals of Glaciology}, 61\penalty0 (82):\penalty0 196--209,
  2020.
\newblock \doi{10.1017/aog.2020.36}.

\bibitem[Li et~al.(2017)Li, Kohout, Doble, Wadhams, Guan, and
  Shen]{li2017rollover}
J.~Li, A.~L. Kohout, M.~J. Doble, P.~Wadhams, C.~Guan, and H.~H. Shen.
\newblock Rollover of apparent wave attenuation in ice covered seas.
\newblock \emph{Journal of Geophysical Research: Oceans}, 122:\penalty0
  8557--8566, 2017.
\newblock \doi{10.1002/2017JC012978}.

\bibitem[Liu et~al.(1992)Liu, Vachon, Peng, and Bhogal]{liu1992wave}
A.~K. Liu, P.~W. Vachon, C.~Y. Peng, and A.~S. Bhogal.
\newblock Wave attenuation in the marginal ice zone during limex.
\newblock \emph{Atmosphere-Ocean}, 30\penalty0 (2):\penalty0 192--206, 1992.
\newblock \doi{10.1080/07055900.1992.9649437}.

\bibitem[Mei et~al.(2005)Mei, Stiassnie, and Yue]{mei2005theory}
Chiang Mei, Michael Stiassnie, and D.Y.-P. Yue.
\newblock \emph{Theory And Applications Of Ocean Surface Waves (Third Edition)
  (In 2 Volumes)}.
\newblock World Scientific, 2005.
\newblock ISBN 978-981-314-717-1.
\newblock \doi{10.1142/10212}.

\bibitem[Mei and Li(2004)]{mei2004evolution}
Chiang~C. Mei and Yile Li.
\newblock Evolution of solitons over a randomly rough seabed.
\newblock \emph{Phys. Rev. E}, 70\penalty0 (1):\penalty0 016302, 2004.
\newblock \doi{10.1103/PhysRevE.70.016302}.

\bibitem[Meylan et~al.(2014)Meylan, Bennetts, and Kohout]{meylan2014in}
M.~H. Meylan, L.~G. Bennetts, and A.~L. Kohout.
\newblock In situ measurements and analysis of ocean waves in the antarctic
  marginal ice zone.
\newblock \emph{Geophysical Research Letters}, 41:\penalty0 5046--5051, 2014.
\newblock \doi{10.1002/2014GL060809}.

\bibitem[Meylan et~al.(2018)Meylan, Bennetts, Mosig, Rogers, Doble, and
  Peter]{meylan2018dispersion}
M.~H. Meylan, L.~G. Bennetts, J.~E.~M. Mosig, W.~E. Rogers, M.~J. Doble, and
  M.~A. Peter.
\newblock Dispersion relations, power laws, and energy loss for waves in the
  marginal ice zone.
\newblock \emph{Journal of Geophysical Research: Oceans}, 123:\penalty0
  3322--3335, 2018.
\newblock \doi{10.1002/2018JC013776}.

\bibitem[Meylan et~al.(2021)Meylan, Horvat, Bitz, and Bennetts]{meylan2021floe}
Michael~H. Meylan, Christopher Horvat, Cecilia~M. Bitz, and Luke~G. Bennetts.
\newblock A floe size dependent scattering model in two- and three-dimensions
  for wave attenuation by ice floes.
\newblock \emph{Ocean Modelling}, 161:\penalty0 101779, 2021.
\newblock ISSN 1463-5003.
\newblock \doi{10.1016/j.ocemod.2021.101779}.

\bibitem[Mokus(2023)]{mokus2023breaking}
Nicolas Guillaume~Alexandre Mokus.
\newblock \emph{Breaking waves in marginal ice zones: numerical study of
  wave-induced sea ice breakup and resulting wave attenuation}.
\newblock Doctor of philosophy - phd, University of Otago, 2023.
\newblock URL \url{https://hdl.handle.net/10523/16330}.

\bibitem[Montiel et~al.(2016)Montiel, Squire, and
  Bennetts]{montiel2016attenuation}
F.~Montiel, V.~A. Squire, and L.~G. Bennetts.
\newblock Attenuation and directional spreading of ocean wave spectra in the
  marginal ice zone.
\newblock \emph{Journal of Fluid Mechanics}, 790:\penalty0 492--522, 2016.
\newblock \doi{10.1017/jfm.2016.21}.

\bibitem[Montiel et~al.(2022)Montiel, Kohout, and Roach]{montiel2022physical}
F.~Montiel, A.~L. Kohout, and L.~A. Roach.
\newblock Physical drivers of ocean wave attenuation in the marginal ice zone.
\newblock \emph{Journal of Physical Oceanography}, 52:\penalty0 889--906, 2022.
\newblock \doi{10.1175/JPO-D-21-0240.1}.

\bibitem[Mosig et~al.(2017)Mosig, Montiel, and Squire]{mosig2017water}
J.~E.~M. Mosig, F.~Montiel, and V.~A. Squire.
\newblock Water wave scattering from a mass loading ice floe of random length
  using generalised polynomial chaos.
\newblock \emph{Wave Motion}, 70:\penalty0 222--239, 2017.
\newblock ISSN 0165-2125.
\newblock \doi{10.1016/j.wavemoti.2016.09.005}.
\newblock Recent Advances on Wave Motion in Fluids and Solids.

\bibitem[Porter(2020)]{porter2020mild}
D.~Porter.
\newblock The mild-slope equations: a unified theory.
\newblock \emph{Journal of Fluid Mechanics}, 887:\penalty0 A29, 2020.
\newblock \doi{10.1017/jfm.2020.21}.

\bibitem[Porter and Porter(2004)]{porter2004approximations}
D.~Porter and R.~Porter.
\newblock Approximations to wave scattering by an ice sheet of variable
  thickness over undulating bed topography.
\newblock \emph{Journal of Fluid Mechanics}, 509:\penalty0 145--179, 2004.
\newblock \doi{10.1017/S0022112004009267}.

\bibitem[Rogers et~al.(2021)Rogers, Meylan, and Kohout]{rogers2021estimates}
W.~Erick Rogers, Michael~H. Meylan, and Alison~L. Kohout.
\newblock Estimates of spectral wave attenuation in antarctic sea ice, using
  model/data inversion.
\newblock \emph{Cold Regions Science and Technology}, 182:\penalty0 103198,
  2021.
\newblock ISSN 0165-232X.
\newblock \doi{10.1016/j.coldregions.2020.103198}.

\bibitem[Squire et~al.(2009)Squire, Vaughan, and Bennetts]{squire2009ocean}
V.~A. Squire, G.~L. Vaughan, and L.~G. Bennetts.
\newblock Ocean surface wave evolvement in the arctic basin.
\newblock \emph{Geophysical Research Letters}, 36:\penalty0 L22502, 2009.
\newblock \doi{10.1029/2009GL040676}.

\bibitem[Thomson et~al.(2021)Thomson, Hošeková, Meylan, Kohout, and
  Kumar]{thomson2021spurious}
J.~Thomson, L.~Hošeková, M.~H. Meylan, A.~L. Kohout, and N.~Kumar.
\newblock Spurious rollover of wave attenuation rates in sea ice caused by
  noise in field measurements.
\newblock \emph{Journal of Geophysical Research: Oceans}, 126:\penalty0
  e2020JC016606, 2021.
\newblock \doi{10.1029/2020JC016606}.

\bibitem[Vaughan et~al.(2009)Vaughan, Bennetts, and Squire]{vaughan2009decay}
Gareth~L. Vaughan, Luke~G. Bennetts, and Vernon~A. Squire.
\newblock The decay of flexural-gravity waves in long sea ice transects.
\newblock \emph{Proceedings of the Royal Society A: Mathematical, Physical and
  Engineering Sciences}, 465\penalty0 (2109):\penalty0 2785--2812, 2009.
\newblock \doi{10.1098/rspa.2009.0187}.

\bibitem[Wadhams and Holt(1991)]{wadhams1991waves}
P.~Wadhams and B.~Holt.
\newblock Waves in frazil and pancake ice and their detection in seasat
  synthetic aperture radar imagery.
\newblock \emph{Journal of Geophysical Research}, 96\penalty0 (C5):\penalty0
  8835--8852, 1991.
\newblock \doi{10.1029/91JC00457}.

\bibitem[Wadhams et~al.(1988)Wadhams, Squire, Goodman, Cowan, and
  Moore]{wadhams1988attenuation}
P.~Wadhams, V.~A. Squire, D.~J. Goodman, A.~M. Cowan, and S.~C. Moore.
\newblock The attenuation rates of ocean waves in the marginal ice zone.
\newblock \emph{Journal of Geophysical Research}, 93\penalty0 (C6):\penalty0
  6799--6818, 1988.
\newblock \doi{10.1029/JC093iC06p06799}.

\bibitem[Wang and Shen(2010)]{wang2010visco}
Ruixue Wang and Hayley~H. Shen.
\newblock Gravity waves propagating into an ice-covered ocean: A viscoelastic
  model.
\newblock \emph{Journal of Geophysical Research: Oceans}, 115\penalty0 (C6),
  2010.
\newblock \doi{10.1029/2009JC005591}.

\bibitem[Weitz and Keller(1950)]{weitz1950reflection}
M.~Weitz and J.~B. Keller.
\newblock Reflection of water waves from floating ice in water of finite depth.
\newblock \emph{Communications on Pure and Applied Mathematics}, 3:\penalty0
  305--318, 1950.
\newblock \doi{10.1002/cpa.3160030306}.

\end{thebibliography}
\end{document}